\documentclass[3p]{elsarticle}

\usepackage{amsmath,amssymb,times,hyperref,doi}

\journal{Physica A}

\biboptions{sort&compress}

\begin{document}

\begin{frontmatter}

\title{Spatial population dynamics: beyond the Kirkwood superposition approximation by advancing to the Fisher-Kopeliovich ansatz}

\author{Igor Omelyan}

\address{Institute for Condensed Matter Physics, National Academy of
Sciences of Ukraine, 1 Svientsitskii Street, UA-79011 Lviv, Ukraine}

\ead{omelyan@icmp.lviv.ua}

\begin{abstract}
The superior Fisher-Kopeliovich closure is applied to the hierarchy of master equations for spatial moments of population dynamics for the first time. As a consequence, the population density, pair and triplet distribution functions are calculated within this closure for a birth-death model with nonlocal dispersal and competition in continuous space. The new results are compared with those obtained by ``exact'' individual-based simulations as well as by the inferior mean-field and Kirkwood superposition approximations. It is shown that the Fisher-Kopeliovich approach significantly improves the quality of the description in a wide range of varying parameters of the model.
\end{abstract}


\end{frontmatter}

\section{Introduction}

Population dynamics (PD) is extensively studied in mathematical biology, ecology, medicine, life and social sciences, etc to predict various effects and phenomena. Many models were proposed during the long history of PD. They include continuum, spatial, continuous-space, lattice, network, individual-based, and other approaches \cite{Murray, Bolker, Gross, Iannelli}. The individual-based models (IBMs) yield the most accurate and detailed information on spatial structure of the system. However, in computer simulations, the IBMs may be numerically very expensive, especially for populations of large sizes \cite{Bolker, South, Lethbridge, Melunis}.

For overcoming drawbacks of existing PD models, over the last two decades there has been an increasing interest in developing spatial moment dynamics (SMD) \cite{BolkerPac, BolkerPaca, Dieckmann, Law, Murrell, Birch, Adams, FinKonK, OvaskFin, Simpson, Plank, Binny, FinKonKoz, Binnya, Surendran, Surendram}. In SMD the populations are described by time-dependent spatial moments (aka reduced distribution \cite{Birch} or correlation \cite{FinKonK, FinKonKoz, Ruelle} functions). The first two of them are local population density and pair correlation function. The SMD approach can be viewed as an extension of the traditional mean-field (MF) theory. The latter is invoked in most PD models to simplify consideration \cite{Morozov}. It should be mentioned that MF totally neglects pair, triplet and higher-order spatial correlations, so that density is the only aggregated parameter. On the other hand, these correlations are explicitly accounted in the SMD theory.

The SMD models were applied to ecological dynamics, surface chemistry reactions, spatial epidemics, herding behaviour, predator-prey metapopulations (see \cite{Bolker, Simpson} and the references therein). These models are particularly useful for detecting patchiness and clustering \cite{Law, Young} in the spatial distribution of different organisms, such as trees in a beech forest \cite{LawIll} or breast cancer cells at an in vitro growth-to-confluence assay \cite{Agnew}. Strictly speaking, the SMD approach allows to reveal subtle effects which are inaccessible for the MF approximation. The former can also be exploited to revisit MF data, e.g., on the formation of patterns in evolution of bacterial colonies \cite{Fuentes}.

The SMD description is exact provided the infinite number of spatial moments are involved in the hierarchy of governing equations \cite{Plank, OvaskFin, OvaskCor, FinKon}. For practical reasons, this hierarchy needs to be truncated (closed) because in computer simulations we cannot operate with the infinite number of moments. Usually the closure is performed for the third-order correlation function, so that numerical solutions are found for the first two equations (which cannot be solved analytically). A lot of truncation schemes \cite{Dieckmann, Murrell, Binny, Binnya} have been used in the literature on PD, including power-1 and -2 closures as well as the power-3 Kirkwood superposition approximation (KSA). It has been established that the closures of higher powers are more accurate than the lower-order ones. The moment truncation is widely used not only in PD but in many other areas when modelling complex systems \cite{Demirel, Kuehn}.

The KSA closure is well known in statistical physics \cite{Kirkwood, Singer, Grouba, Hansen, BenNaim}. Modified and generalized versions of this closure have also been proposed \cite{HernandoG, Attard, Killian, Dufour}. Using the principle of constrained maximum entropy, a corrected KSA has been introduced, too, in the context of PD \cite{Raghib}. A high efficiency of the original KSA closure has been proven not only for spatially homogeneous PD systems in continuous space \cite{Dieckmann, Binny, Binnya, Surendram}, but for homogeneous \cite{Baker, Markham, Markhama} as well as inhomogeneous lattice \cite{Simpsona, Johnston} and off-lattice \cite{Middleton, Matsiaka} models. Quite recently \cite{OmelKoz}, the KSA ansatz has been applied to spatially inhomogeneous birth-death models in continuous space. As a consequence, new subtle effects, possible in real populations, have been discovered.

Despite its indubitable advantage over the MF approximation, in many cases the KSA closure can provide only a qualitative description of the structural behaviour. Being good for the description of segregated states, it performs comparatively poorly in strongly aggregated spatial patterns \cite{Dieckmann, Murrell, Binnya, Raghib}. Some what better results can be achieved by artificially modifying the power-2 closure \cite{Dieckmann, Law, Binnya, Surendram}. However, this modification destroys properties of symmetry for the triplet correlation function and can lead, in general, to unphysical negative values for the density of the system. The origin of mistakes of the KSA closure have been discussed when investigating structural behaviour of populations \cite{Raghib} and thermodynamic properties of liquids \cite{Grouba}.

A natural way to improve the KSA approach is to perform the truncation of the infinite hierarchy of governing equations for spatial moments on the next, higher-order level. In other words, it is necessary to express the quadruplet correlation function in terms of the lower-order spatial moments. Then we come to the power-4 Fisher-Kopeliovich (FK) closure \cite{Singer, Kopeliovich, Somani, Sharkey} originally derived in the theory of liquids more than half a century ago. Despite its obvious superiority over KSA, we know only a few examples on applying the FK closure to actual calculations. For instance, it was used when constructing an entropy equation \cite{Hernando} as well as when studying the structural hydration of biological macromolecules \cite{Hummer, Garcia}. In both cases, enhancement in the description was reported.

Until now, there have been no publications on applications of the FK closure to the field of spatial population dynamics. This can be explained by the fact that the straightforward employment of standard numerical methods for such applications causes heavy computational efforts. The latter increase more than in one order of magnitude with respect to those of KSA, making practically impossible to perform actual SMD/FK calculations.

A key goal of the current work is to determine which level of description for the infinite hierarchy of the spatial-moment master equations is required to obtain quantitative agreement with ''exact'' results of IBM simulations. To accomplish this, in the present paper we propose an efficient numerical algorithm for solving the SMD/FK master equations. This has allowed us to evaluate the local population density, pair and triplet distribution functions for a popular birth-death model with nonlocal dispersal and competition. By direct comparison with the ''exact'' results, a superiority of the power-4 FK ansatz over the power-3 KSA and lower-order approximations is demonstrated.

\section{Theory}

\subsection{General form of the SMD/FK equations}

Consider a population system consisting of $\mathcal{N}(t)=\sum_{a=1}^M \mathcal{N}_a(t)$ interacting agents at time $t$ belonging to $M$ different types and dwelling in the infinite continuous space $\mathbb{R}^d$. Evolution of the system will be described in terms of the first-, second- and third-order spatial moments which are denoted by $F_a^\textrm{(1)}(x,t)$, $F_{ab}^\textrm{(2)}(x,y,t)$ and $F_{abc}^\textrm{(3)}(x,y,z,t)$, respectively. They depend on time $t$ and on the corresponding number of spatial coordinates $x$, $y$, and $z$ of a certain dimensionality $d=1,2$ or $3$, where $x,y,z \in \mathbb{R}^d$ and $a,b,c=1,2,\ldots,M$. The first moment describes the mean density probability of finding a single agent of type $a$ in point $x$ at time $t$. The second moment relates to the finding probability for a pair of agents of types $a$ and $b$ with the location in $x$ and $y$ for the first and second agents, respectively, excluding self-pairs when $a=b$ and $x=y$. Similarly, the third moment $F_{abc}^\textrm{(3)}(x,y,z,t)$ is responsible for the distribution of agents in terms of triplet configurations. Being integrating over the whole space, the spatial moments yield the mean $\langle \ \rangle$ numbers of entities, their pairs, and triplets in the system. In particular, $\int F_a^\textrm{(1)}(x,t) {\rm d} x = \langle \mathcal{N}_a \rangle(t)$ and $\int \int F_{ab}^\textrm{(2)}(x,y,t) {\rm d} x {\rm d} y = \langle \mathcal{N}_a (\mathcal{N}_a-1) \rangle \delta_{ab} + \langle \mathcal{N}_a \mathcal{N}_b \rangle (1-\delta_{ab})$ (see Appendix A, where the moments and mean numbers are rigorously defined).

We will deal with three classes of stochastic events in the population: birth (B), death (D), and movement (M). Then the SMD equations for the first-order spatial moments ($a=1,2,\ldots,M$) can be presented in the form \cite{Plank, Binny, Binnya}:
\begin{equation}
\frac{\partial F_a^\textrm{(1)}(x,t)}{\partial t} = - \sum_{\rm P=D,M} Q_{1,a}^{({\rm P)}}(x,t) F_a(x,t) + \int \sum_{\rm P=B,M} \mu_a^\textrm{(P)}(x,y) Q_{1,a}^\textrm{(P)}(y,t) F_a^\textrm{(1)}(y,t) {\rm d} y \, ,
\end{equation}
where
\begin{equation}
Q_{1,a}^\textrm{(P)}(x,t) = q_a^\textrm{(P)}(x) + \frac{1}{F_a^\textrm{(1)}(x,t)} \sum_{b=1}^M \int w_{ab}^\textrm{(P)}(x,y) F_{ab}^\textrm{(2)}(x,y,t) {\rm d} y
\end{equation}
is the expected rate for event of class ${\rm P}$ in point $x$ at time $t$ concerning an agent of type $a$. This rate contains the intrinsic part $q_a^\textrm{(P)}(x)$ and the neighbour-dependent contribution. The latter appears due to the pair interactions $w_{ab}^\textrm{(P)}(x,y)$ between the agents. The first term in the rhs of Eq.~(1) represents the process of death (\textrm{P$=$D}) of agents in point $x$ at time $t$ or the process of movement of agents (P$=$M) from this point to another location. These two processes lead to the decrease of the local density in $x$. The second term describes spatially integrated process. The first is birth (P$=$B) of an agent at point $y$ complemented by its dispersal to point $x$ with a probability density function $\mu_a^\textrm{(B)}(x,y)$. The second process is movement (P$=$M) of an existing agent from some location $y$ to the current position $x$ drawn from a probability function $\mu_a^\textrm{(M)}(x,y)$. The last two processes result in the increase of the local density in $x$. Note that $w_{ba}^\textrm{(P)}(y,x)=w_{ab}^\textrm{(P)}(x,y)$ and $\mu_a^\textrm{(P)}(y,x)=\mu_a^\textrm{(P)}(x,y)$.

The birth dispersal and movement dispersal functions are normalized so that $\int \mu_a^\textrm{(B,M)} (x,y) {\rm d} y = 1$. This means that a born agent can be located only in one point of coordinate space. Also, this guarantees that the sum of direct (from $x$) and opposite (to $x$) movements does not change the total number of agents of a given type. The dispersal functions $\mu_a^\textrm{(B,M)}(x,y)$ are assumed to be independent of the types and locations of other agents in space. The interaction kernels are not necessarily to be normalized on 1, so that $\int w_{ab}^\textrm{(B,D,M)}(x,y) {\rm d} y=c_{ab}^\textrm{(B,D,M)}(x)$, where $c_{ab}^\textrm{(B,D,M)}(x)$ are the interaction strength parameters. These kernels are related to the neighbour-dependent birth (sexual reproduction), death (due to competition), and movement (collective motion). The intrinsic birth, death, and movement rates $q_a^\textrm{(B,D,M)}(x)$ account for vegetative reproduction, endogenic mortality, and own motility, respectively.

The SMD equations for the second-order moments ($a,b=1,2,\ldots,M$) are \cite{Plank, Binny, Binnya}:
\begin{align}
\frac{\partial F_{ab}^\textrm{(2)}(x,y,t)}{\partial t} =&\ - \sum_{\rm P=D,M} Q_{2,ab}^\textrm{(P)}(x,y,t) F_{ab}^\textrm{(2)}(x,y,t) + \int \sum_{\rm P=B,M} \mu_a^\textrm{(P)}(z,x) Q_{2,ab}^\textrm{(P)}(z,y,t) F_{ab}^\textrm{(2)}(z,y,t) {\rm d} z \nonumber \\[5pt] &\ + \mu_b^\textrm{(B)}(y,x) Q_{1,b}^\textrm{(B)}(y,t) F_b^\textrm{(1)}(y,t) \delta_{ab} + [ a,b,x,y \to b,a,y,x ] \, ,
\end{align}
where
\begin{equation}
Q_{2,ab}^\textrm{(P)}(x,y,t) = q_a^\textrm{(P)}(x) + \frac{1}{F_{ab}^\textrm{(2)}(x,y,t)} \sum_{c=1}^M \int w_{ac}^\textrm{(P)}(x,z) F_{abc}^\textrm{(3)}(x,y,z,t) {\rm d} z + w_{ab}^\textrm{(P)}(x,y)
\end{equation}
is the expected rate for event of class ${\rm P}$ at time $t$ concerning a pair of agents of types $a$ and $b$ located in $x$ and $y$. As in Eq.~(2), this rate consists of the intrinsic and neighbour-dependent parts. The extra third term appears here because the third-order function $F_{abc}^\textrm{(3)}(x,y,z,t)$ excludes triplets containing self-pairs, so that the integral term in Eq.~(4) only measures the contribution of third-party agents, distinct from the pair of agents in $x$ and $y$. Therefore, the effect of the agent in $y$ on the focal agent in $x$ must be added as a separate term, $w_{ab}^\textrm{(P)}(x,y)$. The first two terms in the rhs of Eq.~(3) represent the same processes as those of Eq.~(1) for the first-order distribution function. The third term has been included to cover the case in which a pair is created by the agent at $y$ giving birth to a new agent at $x$. The Kronecker delta $\delta_{ab}$ stipulates that this can only happen if the two agents are of the same type (i.e. $a = b$). No such term is needed for movement of the agent from $y$ to $x$ as this event does not create a pair. Finally, the last (fourth) term means that all the three previous terms should be repeated using the substitution $[a,b,x,y \to b,a,y,x]$. This is necessary to satisfy the symmetric property $F_{ba}^\textrm{(2)}(y,x,t) = F_{ab}^\textrm{(2)}(x,y,t)$. Note that the second- and first-order spatial moments are connected by the integral relations $\int F_{aa}^\textrm{(2)}(x,y,t) {\rm d} y = (\langle \mathcal{N}_a^2 \rangle / \langle \mathcal{N}_a \rangle - 1) F_{aa}^\textrm{(1)}(x,t)$ and $\sum_{b \ne a} \int F_{ab}^\textrm{(2)}(x,y,t) {\rm d} y = \sum_{b \ne a} \langle \mathcal{N}_a \mathcal{N}_b \rangle / \langle \mathcal{N}_a \rangle F_a^\textrm{(1)}(x,t)$. Here $\langle \mathcal{N}_a \rangle \equiv \langle \mathcal{N}_a(t) \rangle \equiv \langle \mathcal{N}_a \rangle(t)$ and $\langle \mathcal{N}_a^2 \rangle \equiv \langle \mathcal{N}_a^2 \rangle(t)$, so that time variable $t$ was omitted in the relations to simplify their presentation.

The SMD equations for the third-order spatial moments ($a,b,c=1,2,\ldots,M$) can be cast as \cite{Plank, Binny, Binnya}:
\begin{align}
\frac{\partial F_{abc}^\textrm{(3)}(x,y,z,t)}{\partial t} =&\ - \ Q_{3,abc}^\textrm{(D,M)}(x,y,z,t) F_{abc}^\textrm{(3)}(x,y,z,t) 
+ \int \mu_a^\textrm{(B,M)}(u,x) Q_{3,abc}^\textrm{(P)}(u,y,z,t) F_{abc}^\textrm{(3)}(u,y,z,t) {\rm d} u \nonumber \\ &\ + \left( \mu_b^\textrm{(B)}(y,x) Q_{2,bc}^\textrm{(B)}(y,z,t) \delta_{ab} + \mu_c^\textrm{(B)}(z,x) Q_{2,cb}^\textrm{(B)}(z,y,t) \delta_{ac} \right) F_{bc}^\textrm{(2)}(y,z,t) \nonumber \\[4pt] &\ + [ a,b,c,x,y,z \to b,a,c,y,x,z ] + [ a,b,c,x,y,z \to c,a,b,z,x,y ] \, ,
\end{align}
where
\begin{equation}
Q_{3,abc}^\textrm{(P)}(x,y,z,t) = q_a^\textrm{(P)}(x) + \frac{1}{{F_{abc}^\textrm{(3)}(x,y,z,t)}} \sum_{d=1}^M \int w_{ad}^\textrm{(P)}(x,u) F_{abcd}^\textrm{(4)}(x,y,z,u,t) {\rm d} u + w_{ab}^\textrm{(P)}(x,y) + w_{ac}^\textrm{(P)}(x,z)
\end{equation}
is the event rate for triplet spatial configurations. The effect of neighbour agents in $y$ and $z$ on the focal agent in $x$ is included by the last two separate terms in the rhs of Eq.~(6). We see that Eqs.~(5) and (6) are a straightforward extension of Eqs.~(3) and (4) for the second-order function. They contain a dependence on the moment of the fourth order $F_{abcd}^\textrm{(4)}(x,y,z,u,t)$ which is the density of quadruplets in the rate equation (6). The two terms in the second line of the rhs of Eq.~(5) describe a creation of a triplet at $x$ caused by births from parents at $y$ and $z$, and with all events repeated at $y$ and $z$, as indicated by permutations in the square brackets.

Now we rewrite the equations (1), (3), and (5) for the first-, second-, and third-order spatial moments more explicitly by directly substituting expressions Eqs.~(2), (4), and (6) into them. Then one finds
\begin{align}
& \frac{\partial F_a^\textrm{(1)}(x,t)}{\partial t} = - q_a^\textrm{(D,M)}(x) F_a^\textrm{(1)}(x,t) - \sum_{b=1}^M \int w_{ab}^\textrm{(D,M)}(x,y) F_{ab}^\textrm{(2)}(x,y,t) {\rm d} y \nonumber \\[-2pt] & + \int \mu_a^\textrm{(B,M)}(x,y) \sum_{b=1}^M \int w_{ab}^\textrm{(B,M)}(y,z) F_{ab}^\textrm{(2)}(y,z,t) {\rm d} z {\rm d} y + \int \mu_a^\textrm{(B,M)}(x,y) q_a^\textrm{(B,M)}(y) F_a^\textrm{(1)}(y,t) {\rm d} y \, ,
\end{align}
\begin{align}
& \frac{\partial F_{ab}^\textrm{(2)}(x,y,t)}{\partial t} = - \left( q_a^\textrm{(D,M)}(x) + w_{ab}^\textrm{(D,M)}(x,y) \right) F_{ab}^\textrm{(2)}(x,y,t) - \sum_{c=1}^M \int w_{ac}^\textrm{(D,M)}(x,z) F_{abc}^\textrm{(3)}(x,y,z,t) {\rm d} z \nonumber \\[-2pt] & + \int \mu_a^\textrm{(B,M)}(z,x) \left( \left( q_a^\textrm{(B,M)}(z) + w_{ab}^\textrm{(B,M)}(z,y) \right) F_{ab}^\textrm{(2)}(z,y,t) + \sum_{c=1}^M \int w_{ac}^\textrm{(B,M)}(z,u) F_{abc}^\textrm{(3)}(z,y,u,t) {\rm d} u \right) {\rm d} z \nonumber \\ & + \delta_{ab} \mu_b^\textrm{(B)}(y,x) \left( q_b^\textrm{(B)}(y) F_b^\textrm{(1)}(y,t) + \sum_c \int w_{bc}^\textrm{(B)}(y,z) F_{bc}^\textrm{(2)}(y,z,t) {\rm d} z \right) + [a,b,x,y \to b,a,y,x] \, ,
\end{align}
and
\begin{align}
& \frac{\partial F_{abc}^\textrm{(3)}(x,y,z,t)}{\partial t} = - \left( q_a^\textrm{(D,M)}(x) + w_{ab}^\textrm{(D,M)}(x,y) + w_{ac}^\textrm{(D,M)}(x,z) \right) F_{abc}^\textrm{(3)}(x,y,z,t) \nonumber \\[5pt] & + \int \mu_a^\textrm{(B,M)}(u,x) \left( q_a^\textrm{(B,M)}(u) + w_{ab}^\textrm{(B,M)}(u,y) + w_{ac}^\textrm{(B,M)}(u,z) \right) F_{abc}^\textrm{(3)}(u,y,z,t) {\rm d} u \nonumber \\ & + \delta_{ab} \mu_b^\textrm{(B)}(y,x) \left( \left( q_b^\textrm{(B)}(y) + w_{bc}^\textrm{(B)}(y,z) \right) F_{bc}^\textrm{(2)}(y,z,t) + \sum_{d=1}^M \int w_{bd}^\textrm{(B)}(y,u) F_{bcd}^\textrm{(3)}(y,z,u,t) {\rm d} u \right) \nonumber \\ & + \delta_{ac} \mu_c^\textrm{(B)}(z,x) \left( \left( q_c^\textrm{(B)}(z) + w_{cb}^\textrm{(B)}(z,y) \right) F_{bc}^\textrm{(2)}(y,z,t) + \sum_{d=1}^M \int w_{cd}^\textrm{(B)}(z,u) F_{cbd}^\textrm{(3)}(z,y,u,t) {\rm d} u \right) \nonumber \\[1pt] & - \sum_{d=1}^M \int w_{ad}^\textrm{(D,M)}(x,u) F_{abcd}^\textrm{(4)}(x,y,z,u,t) {\rm d} u + \int \mu_a^\textrm{(B,M)}(u,x) \sum_{d=1}^M \int w_{ad}^\textrm{(B,M)}(u,u') F_{abcd}^\textrm{(4)}(u,y,z,u',t) {\rm d} u' {\rm d} u \nonumber \\[8pt] & + [a,b,c,x,y,z \to b,a,c,y,x,z] + [a,b,c,x,y,z \to c,a,b,z,x,y] \, ,
\end{align}
where notations (D,M) or (B,M) imply that both terms with (D) and (M) or (B) and (M) are included. As can be seen from Eqs.~(7)--(9), the equation for the $p$th-order spatial moment ($p=1,2,3$) requires the knowledge of the moment of the next $(p+1)$th order. Analogously, the equations for spatial moments of arbitrary higher orders can be constructed, resulting in the infinity hierarchy Ref.~\cite{Plank}. It is very similar to the Bogoliubov (or Bogolyubov-Born-Green-Kirkwood-Yvon) chain of equations \cite{Dobrushin, Bogoliubov} for correlation functions in the Hamiltonian dynamics of continuum systems of interacting particles.

Since we deal with the description on the third ($p=3$) level, the governing equations (7)--(9) should be complemented by the closure relation for the fourth-order moment. According to FK this can be done using the following relation \cite{Singer, Kopeliovich, Somani, Sharkey}:
\begin{equation}
F_{abcd}^\textrm{(4)}(x,y,z,u,t) = \frac{F_{abc}^\textrm{(3)}(x,y,z,t) F_{abd}^\textrm{(3)}(x,y,u,t) F_{acd}^\textrm{(3)}(x,z,u,t) F_{bcd}^\textrm{(3)}(y,z,u,t)}{\displaystyle\frac{F_{ab}^\textrm{(2)}(x,y,t) F_{ac}^\textrm{(2)}(x,z,t) F_{ad}^\textrm{(2)}(x,u,t) F_{bc}^\textrm{(2)}(y,z,t) F_{bd}^\textrm{(2)}(y,u,t) F_{cd}^\textrm{(2)}(z,u,t)}{F_a^\textrm{(1)}(x,t) F_b^\textrm{(1)}(y,t) F_c^\textrm{(1)}(z,t) F_d^\textrm{(1)}(u,t)}} \, ,
\end{equation}
where the fourth-order correlation function is expressed in terms of the lower-order ones.

In view of the closure relation (10), equations (7)--(9) constitute a very complicated set of $M+M^2+M^3$ coupled partial integro-differential equations with respect to the same number of unknown functions, $F_a^\textrm{(1)}(x,t)$, $F_{ab}^\textrm{(2)}(x,y,t)$, and $F_{abc}^\textrm{(3)}(x,y,z,t)$. Taking into account the symmetry properties $F_{ba}^\textrm{(2)}(y,x,t) = F_{ab}^\textrm{(2)}(x,y,t)$ and $F_{abc}^\textrm{(3)}(x,y,z,t) = F_{acb}^\textrm{(3)}(x,z,y,t) = F_{bac}^\textrm{(3)}(y,x,z,t) = F_{bca}^\textrm{(3)}(y,z,x,t) = F_{cab}^\textrm{(3)}(z,x,y,t) = F_{cba}^\textrm{(3)}(z,y,x)$, where $a,b,c=1,2,\ldots,M$, the total number of equations and independent functions can be reduced from $M+M^2+M^3$ to $M+(M+1)M/2+(M+2)(M+1)M/6$. These equations have never been solved before that can be explained by the fact that huge difficulties arise immediately when trying to handle them by existing numerical methods.

\subsection{Spatially homogeneous limit}

Because the SMD master equations (7)--(9) in its most general representation are too complicated, we consider the limit of spatially homogeneity. This is a common practice which is used in most papers \cite{BolkerPac, Dieckmann, Law, Murrell, Adams, Binny, Binnya} on spatial PD. The homogeneity does not preclude spatial structure (i.e. departures from a spatial Poisson process): the agents can generate it themselves. Although agent density is spatially uniform on averaging over many independent realisations, strong spatial correlations (such as clustering or disaggregation) can still appear due to the neighbour-dependent birth, death and movements, or the correlation between the locations of parent and offspring.

In the homogeneity limit, the intrinsic and density functions do not change on coordinate $x$, i.e., $q_a^\textrm{(P)}(x) \equiv q_a^\textrm{(P)}$ and $F_a^\textrm{(1)}(x,t) \equiv F_a^\textrm{(1)}(t)$, while the second-order moment will depend only on the difference $\xi=y-x$ (distance between entities) and not on $x$ and $y$ separately, i.e., $F_{ab}^\textrm{(2)}(x,y,t) \equiv F_{ab}^\textrm{(2)}(\xi,t)$. Moreover, $F_{ab}^\textrm{(2)}(\xi,t) \equiv F_{ab}^\textrm{(2)}(|\xi|,t)$. The same concerns the interaction and dispersal kernels, so that $w_{ab}^\textrm{(P)}(x,y) \equiv w_{ab}^\textrm{(P)}(|x-y|)$ and $\mu_a^\textrm{(P)}(x,y) \equiv \mu_{ab}^\textrm{(P)}(|x-y|)$. Similarly, the third-order moment $F_{abc}^\textrm{(3)}(x,y,z,t)$ being originally dependent on three spatial coordinates ($x,y,z$) now will be a function of two variables, i.e., $F_{abc}^\textrm{(3)}(\xi,\xi',t)$, where $\xi'=z-x$.

Therefore, making the formal replacements $y-x=\xi$ and ${\rm d} y={\rm d} \xi$, Eq.~(7) for the density functions transforms to
\begin{equation}
\frac{{\rm d} F_a^\textrm{(1)}(t)}{{\rm d} t} = \Big(q_a^\textrm{(B)} - q_a^\textrm{(D)} \Big) F_a^\textrm{(1)}(t) + \sum_b \int \Big( w_{ab}^\textrm{(B)}(\xi) - w_{ab}^\textrm{(D)}(\xi) \Big) F_{ab}^\textrm{(2)}(\xi,t) {\rm d} \xi \, .
\end{equation}
Further, making the replacements $y-x=\xi$, $z-x=\xi'$, $z-y=\xi'-\xi$, and ${\rm d} z = {\rm d} \xi'$, Eq.~(8) for the second-order spatial moments takes the form
\begin{eqnarray}
&& \frac{\partial F_{ab}^\textrm{(2)}(\xi,t)}{\partial t} = - \left( q_a^\textrm{(D,M)} + w_{ab}^\textrm{(D,M)}(\xi) \right) F_{ab}^\textrm{(2)}(\xi,t) - \sum_c \int w_{ac}^\textrm{(D,M)}(\xi') F_{abc}^\textrm{(3)}(\xi,\xi',t) {\rm d} \xi' \nonumber \\[6pt] && + \int \mu_a^\textrm{(B,M)}(\xi') \Big( \big( q_a^\textrm{(B,M)} + w_{ab}^\textrm{(B,M)}(\xi'-\xi) \big) F_{ab}^\textrm{(2)}(\xi'-\xi,t) + \sum_c \int w_{ac}^\textrm{(B,M)}(\xi'') F_{abc}^\textrm{(3)}(\xi'-\xi,\xi'',t) {\rm d} \xi'' \Big) {\rm d} \xi' \nonumber \\[5pt] && + \ \mu_b^\textrm{(B)}(-\xi) \Big( q_b^\textrm{(B)} F_b^\textrm{(1)}(t) + \sum_c \int w_{bc}^\textrm{(P)}(\xi') F_{bc}^\textrm{(2)}(\xi',t) {\rm d} \xi' \Big) \delta_{ab} + [a,b,\xi \to b,a,-\xi] \, .
\end{eqnarray}
In the new variables, the symmetry properties of the second- and third-order distribution functions read: $F_{ab}^\textrm{(2)}(\xi,t)=F_{ba}^\textrm{(2)}(-\xi,t)$ as well as $F_{abc}^\textrm{(3)}(\xi,\xi',t) = F_{acb}^\textrm{(3)}(\xi',\xi,t) = F_{bac}^\textrm{(3)}(-\xi,\xi'-\xi,t) = F_{bca}^\textrm{(3)}(\xi'-\xi,-\xi,t) = F_{cab}^\textrm{(3)}(-\xi',\xi-\xi',t) = F_{cba}^\textrm{(3)}(\xi-\xi',-\xi',t)$. Note also that $w_{ab}^\textrm{(P)}(-\xi) = w_{ab}^\textrm{(P)}(\xi)$ and $\mu_a^\textrm{(P)}(-\xi) = \mu_a^\textrm{(P)}(\xi)$. The last term in the rhs of Eq.~(12) means that all the previous terms should be repeated using the substitution $a,b,\xi \to b,a,-\xi$. This is necessary to satisfy the above symmetric property $F_{ba}^\textrm{(2)}(-\xi,t) = F_{ab}^\textrm{(2)}(\xi,t)$.

Consider now the most complicated expression given by Eq.~(9) for the third-order spatial moments. Introducing the new variables $y-x=\xi$, $z-x=\xi'$, $z-y=\xi'-\xi$, $u-x=\xi''$, $z-u=x+\xi'-u=\xi'-\xi''$, $y-u=x+\xi-u=\xi-\xi''$ with ${\rm d} u = {\rm d} \xi''$, these expression modifies to
\begin{eqnarray}
&& \frac{\partial F_{abc}^\textrm{(3)}(\xi,\xi',t)}{\partial t} = - \ \left( q_a^\textrm{(D,M)} + w_{ab}^\textrm{(D,M)}(\xi) + w_{ac}^\textrm{(D,M)}(\xi') \right) F_{abc}^\textrm{(3)}(\xi,\xi',t) \nonumber \\[4pt] && + \int \mu_a^\textrm{(B,M)}(-\xi'')
\Big( q_a^\textrm{(B,M)} + \ w_{ab}^\textrm{(B,M)}(\xi-\xi'') + w_{ac}^\textrm{(B,M)}(\xi'-\xi'') \Big) F_{abc}^\textrm{(3)}(\xi-\xi'',\xi'-\xi'',t) {\rm d} \xi'' \nonumber \\[2pt] && + \ \delta_{ab} \mu_b^\textrm{(B)}(-\xi) \left( \Big( q_b^\textrm{(B)} + w_{bc}^\textrm{(B)}(\xi'-\xi) \Big) F_{bc}^\textrm{(2)}(\xi'-\xi,t) + \sum_d \int w_{ac}^\textrm{(B)}(\xi'') F_{bcd}^\textrm{(3)}(\xi'-\xi,\xi'',t) {\rm d} \xi'' \right) \nonumber \\ && + \ \delta_{ac} \mu_c^\textrm{(B)}(-\xi') \left( \Big( q_c^\textrm{(B)} + w_{cb}^\textrm{(B)}(\xi-\xi') \Big) F_{bc}^\textrm{(2)}(\xi'-\xi,t) + \sum_d \int w_{ac}^\textrm{(B)}(\xi'') F_{cbd}^\textrm{(3)}(\xi-\xi',\xi'',t) {\rm d} \xi'' \right) \nonumber \\[4pt] && + \int \mu_a^\textrm{(B,M)}(-\xi'') \sum_d \int w_{ad}^\textrm{(B,M)}(\xi''') F_{abcd}^{(4)}(\xi-\xi'',\xi'-\xi'',\xi''',t) {\rm d} \xi'''
{\rm d} \xi'' \nonumber \\[2pt] && - \sum_d \int w_{ad}^\textrm{(D,M)}(\xi'') F_{abcd}^{(4)}(\xi,\xi',\xi'',t) {\rm d} \xi'' + [a,b,c,\xi,\xi' \to b,a,c,-\xi,\xi'-\xi] \nonumber \\[6pt] && + \ [a,b,c,\xi,\xi' \to c,a,b,-\xi',\xi-\xi'] \, ,
\end{eqnarray}
where the last two terms mean that all the previous terms should be repeated twice using the corresponding substitutions to satisfy the mentioned above symmetric properties of $F_{abc}^\textrm{(3)}(\xi,\xi',t)$.

The FK ansatz (10) in the spatially homogeneous space reads as
\begin{eqnarray}
F_{abcd}^{(4)}(\xi,\xi',\xi'',t) = \frac{F_{abc}^\textrm{(3)}(\xi,\xi',t) F_{abd}^\textrm{(3)}(\xi,\xi'',t) F_{acd}^\textrm{(3)}(\xi',\xi'',t) F_{bcd}^\textrm{(3)}(\xi'-\xi,\xi''-\xi,t)}{\displaystyle\frac{F_{ab}^\textrm{(2)}(\xi,t) F_{ac}^\textrm{(2)}(\xi',t) F_{ad}^\textrm{(2)}(\xi'',t) F_{bc}^\textrm{(2)}(\xi'-\xi,t) F_{bd}^\textrm{(2)}(\xi''-\xi,t) F_{cd}^\textrm{(2)}(\xi''-\xi',t)}{F_a^\textrm{(1)}(t) F_b^\textrm{(1)}(t) F_c^\textrm{(1)}(t) F_d^\textrm{(1)}(t)}} \, .
\end{eqnarray}

Again as in the general case [Eqs.~(7)--(9)], the SMD equations (11)--(13) complemented by the FK closure (14) constitute a system of $M+(M+1)M/2+(M+2)(M+1)M/6$ coupled partial integro-differential equations with respect to the same number of unknown functions, $F_a^\textrm{(1)}(t)$, $F_{ab}^\textrm{(2)}(\xi,t)$, and $F_{abc}^\textrm{(3)}(\xi,\xi',t)$. But now the maximal number of spatial variables decreases from three to two, opening possibilities to solve these equations numerically.

\subsection{Dispersal and competition kernels}

The dependencies of the interaction and dispersal kernels on distance between entites are usually chosen in the forms of the Gaussians
\begin{equation}
w_{ab}^\textrm{(P)}(\xi) = \frac{c_{ab}^\textrm{(P)}}{\big(2 \pi \sigma_{ab}^\textrm{(P)}\big)^{\frac{d}{2}}} \exp \left(-\frac{\xi^2}{2 {\sigma_{ab}^\textrm{(P)}}^2}\right) \, , \ \ \ \ \ \ \ \ \mu_a^\textrm{(P)}(\xi) = \frac{1}{(2 \pi s_{ab}^\textrm{(P)})^{\frac{d}{2}}} \exp \left(-\frac{\xi^2}{2 {s_{ab}^\textrm{(P)}}^2}\right) \, ,
\end{equation}
or the top-hat (Heaviside-like) functions
\begin{equation}
w_{ab}^\textrm{(P)}(\xi) = \left\{
\begin{array}{cc}
\displaystyle \frac{c_{ab}^\textrm{(P)}}{\varrho_d {\sigma_{ab}^\textrm{(P)}}^d} \, , & |\xi| \le \sigma_{ab}^\textrm{(P)} \\ 0 \, ,& |\xi| > \sigma_{ab}^\textrm{(P)}
\end{array} \right. , \ \ \ \ \ \ \ \ \mu_a^\textrm{(P)}(\xi) = \left\{
\begin{array}{cc}
\displaystyle \frac{1}{\varrho_d {s_{ab}^\textrm{(P)}}^d} \, , & |\xi| \le s_{ab}^\textrm{(P)} \\ 0 \, ,& |\xi| > s_{ab}^\textrm{(P)}
\end{array} \right. .
\end{equation}
Here $c_{ab}^\textrm{(P)}$ and $\sigma_{ab}^\textrm{(P)}$ are the intensities and ranges of the interactions, respectively, while $s_{ab}^\textrm{(P)}$ denote the ranges of the dispersal functions. In the case of the top-hat functions, the truncation at $d \ge 2$ can be carried out using square or cube areas, so that $\varrho_d = 2^d$. Alternatively we can use circle or sphere areas for which $\varrho_d = 2$, $\pi$, or $4 \pi/3$ at $d=1,2$, or $3$. Then the normalizations $\int w_{ab}^\textrm{(P)}(\xi) {\rm d} \xi = c_{ab}^\textrm{(P)}$ and $\int \mu_a^\textrm{(P)}(\xi) {\rm d} \xi = 1$ will be satisfied.

In view of the above, we have, in general, up 11 independent parameters of the SMD model, namely, $q_a^\textrm{(B)}$, $q_a^\textrm{(D)}$, $q_a^\textrm{(M)}$, $c_{ab}^\textrm{(B)}$, $c_{ab}^\textrm{(D)}$, $c_{ab}^\textrm{(M)}$, $\sigma_{ab}^\textrm{(B)}$, $\sigma_{ab}^\textrm{(D)}$, $\sigma_{ab}^\textrm{(M)}$, $s_{ab}^\textrm{(B)}$, $s_{ab}^\textrm{(M)}$, even for a one-component population ($a=b=M=1$), and up 30 such parameters at $M=2$. For $M > 2$, this number can achieve much larger values.

\section{Numerical algorithm}

\subsection{Spatial discretization}

In order to solve the SMD/FK equations (11)--(14) by computer calculations it is necessary, first of all, to perform their discretization in coordinate space. Let $\xi_i$ be the grid points of $\xi \in \mathbb{R}^d$ uniformly distributed over the area $[-L,L]^d$ at mesh $h=(L/N)^d$, where $i=0,1,2,\ldots,N^d$ with $\xi_0=0 < \xi_1 < \ldots < \xi_N=L$ and $\xi_{-i}=-\xi_i$ along each dimension $d$. This area presents an interval, a square, or a cube in the cases $d=1$, $2$, or $3$, respectively. Then the discrete counterparts of Eqs.~(11)--(13) are of the forms
\begin{equation}
\frac{{\rm d} F_a^\textrm{(1)}}{{\rm d} t} = \Big(q_a^\textrm{(B)} - q_a^\textrm{(D)} \Big) F_a^\textrm{(1)} + h \sum_{b,i} \Big( w_{ab,i}^\textrm{(B)} - w_{ab,i}^\textrm{(D)} \Big) \zeta_i F_{ab,i}^\textrm{(2)} \, ,
\end{equation}
\begin{align}
\frac{{\rm d} F_{ab,i}^\textrm{(2)}}{{\rm d} t} =&\ - \left( q_a^\textrm{(D,M)} + w_{ab,i}^\textrm{(D,M)} \right) F_{ab,i}^\textrm{(2)} - h \sum_{c,j} w_{ac,j}^\textrm{(D,M)} \zeta_j F_{abc,ij}^\textrm{(3)} \nonumber \\ &\ + \ h \sum_j \mu_{a,j}^\textrm{(B,M)} \big( q_a^\textrm{(B,M)} + w_{ab,j-i}^\textrm{(B,M)} \big) \zeta_j F_{ab,j-i}^\textrm{(2)} + h^2 \sum_j \mu_{a,j}^\textrm{(B,M)} \zeta_j \sum_{c,k} w_{ac,k}^\textrm{(B,M)} \zeta_k F_{abc,j-i,k}^\textrm{(3)} \nonumber \\[5pt] &\ + \ \mu_{b,-i}^\textrm{(B)} \Big( q_b^\textrm{(B)} F_b^\textrm{(1)} + h \sum_{c,j} w_{bc,j}^\textrm{(B)} \zeta_j F_{bc,j}^\textrm{(2)} \Big) \delta_{ab} + [a,b,i \to b,a,-i] \, ,
\end{align}
and
\begin{align}
\frac{{\rm d} F_{abc,ij}^\textrm{(3)}}{{\rm d} t} =&\ - \ \left( q_a^\textrm{(D,M)} + w_{ab,i}^\textrm{(D,M)} + w_{ac,j}^\textrm{(D,M)} \right) F_{abc,ij}^\textrm{(3)} \nonumber \\[4pt] &\ + \ h \sum_k \zeta_{-k} \mu_{a,-k}^\textrm{(B,M)} \Big( q_a^\textrm{(B,M)} + w_{ab,i-k}^\textrm{(B,M)} + w_{ac,j-k}^\textrm{(B,M)} \Big) F_{abc,i-k,j-k}^\textrm{(3)} \nonumber \\ &\ + \ \delta_{ab} \mu_{b,-i}^\textrm{(B)} \left( \Big( q_b^\textrm{(B)} + w_{bc,j-i}^\textrm{(B)} \Big) F_{bc,j-i}^\textrm{(2)} + h \sum_{d,k} w_{ac,k}^\textrm{(B)} \zeta_k F_{bcd,j-i,k}^\textrm{(3)} \right) \nonumber \\ &\ + \ \delta_{ac} \mu_{c,-j}^\textrm{(B)} \left( \Big( q_c^\textrm{(B)} + w_{cb,i-j}^\textrm{(B)} \Big) F_{bc,j-i}^\textrm{(2)} + h \sum_{d,k} w_{ac,k}^\textrm{(B)} \zeta_k F_{cbd,i-j,k}^\textrm{(3)} \right) \nonumber \\[4pt] &\ + \ h^2 \sum_k \mu_{a,-k}^\textrm{(B,M)} \zeta_{-k} \sum_{d,l} w_{ad,l}^\textrm{(B,M)} \zeta_l F_{abcd,i-k,j-k,l}^{(4)} - h \sum_{d,k} w_{ad,k}^\textrm{(D,M)} \zeta_k F_{abcd,ijk}^{(4)} \nonumber \\[6pt] &\ + \ [a,b,c,i,j \to b,a,c,-i,j-i] + [a,b,c,i,j \to c,a,b,-j,i-j] \, .
\end{align}
with $i,j=-N^d,-N^d+1,\ldots,0,1,2,\ldots,N^d$. The disretized version of the KF closure (14) is
\begin{align}
F_{abcd,ijk}^{(4)} = \frac{F_{abc,ij}^\textrm{(3)} F_{abd,ik}^\textrm{(3)} F_{acd,jk}^\textrm{(3)} F_{bcd,j-i,k-i}^\textrm{(3)}}{\displaystyle\frac{F_{ab,i}^\textrm{(2)} F_{ac,j}^\textrm{(2)} F_{ad,k}^\textrm{(2)} F_{bc,j-i}^\textrm{(2)} F_{bd,k-i}^\textrm{(2)} F_{cd,k-j}^\textrm{(2)}}{F_a^\textrm{(1)}(t) F_b^\textrm{(1)}(t) F_c^\textrm{(1)}(t) F_d^\textrm{(1)}(t)}} \, .
\end{align}

In Eqs.~(17)--(19), the sums over $i$, $j$, $k$, and $l$ (from $-N^d$ to $N^d$) represent the spatial integrals over $\xi$, $\xi'$, $\xi''$, and $\xi'''$, while $F_{ab,i}^\textrm{(2)} = F_{ab}^\textrm{(2)}(\xi_i,t)$ and $F_{abc,ij}^\textrm{(3)} = F_{abc}^\textrm{(3)}(\xi_i,\xi_j,t)$ are the values of the second- and third-order moments in the grid points. The interaction and dispersal kernels are discretized similarly, so that $w_{ab,i}^\textrm{(P)} = w_{ab}^\textrm{(P)}(\xi_i)$ and $\mu_{a,i}^\textrm{(P)} = \mu_a^\textrm{(P)}(\xi_i)$. The weights $\zeta_i=\zeta_{-i}$ were introduced to improve precision of the numerical integration over coordinate space. They are determined according to the chosen method (composition Simpson, trapezoidal one or others, see subsection~\textit{3.2}). Remember also that $F_{ab,i}^\textrm{(2)}=F_{ba,-i}^\textrm{(2)}$ and $F_{abc,ij}^\textrm{(3)} = F_{acb,ji}^\textrm{(3)} = F_{bac,-i,j-i}^\textrm{(3)} = F_{bca,j-i,-i}^\textrm{(3)} = F_{cab,-j,i=j}^\textrm{(3)} = F_{cba,i-j,-j}^\textrm{(3)}$ as well as $w_{ba,-i}^\textrm{(P)} = w_{ab,i}^\textrm{(P)}$ and $\mu_{a,-i}^\textrm{(P)} = \mu_{a,i}^\textrm{(P)}$ owing to the symmetry properties. In addition, $F_{ab,i}^\textrm{(2)} \equiv F_{ab,|i|}^\textrm{(2)}$ and $F_{ba,i}^\textrm{(2)} = F_{ab,i}^\textrm{(2)}$. Analogously we have for the interaction and dispersal kernels that $w_{ab,i} \equiv w_{ab,|i|}$ with $w_{ba,i}=w_{ab,i}$ and $\mu_{a,i} \equiv \mu_{a,|i|}$.

The basic length $L$ should be sufficiently long with respect to all characteristic coordinate scales of the population system in order to exclude boundary effects. In particular, at $|\xi| > L$ the values of the interaction and dispersal kernels should be negligibly small, i.e, $w_{ab}^\textrm{(P)}(\xi) \approx 0$ and $\mu_{a}^\textrm{(P)}(\xi) \approx 0$, that is possible provided $L \gg \max(\sigma_{ab}^\textrm{(P)}, s_a^\textrm{(P)}\big)$. Yet, at this length, the second- and third-order spatial moments must exhibit their asymptotic (uncorrelated MF-like) behaviour, $F_{ab}^\textrm{(2)}(\xi,t) \approx F_a^\textrm{(1)}(t) F_b^\textrm{(1)}(t)$ and $F_{abc}^\textrm{(2)}(\xi,t) \approx F_a^\textrm{(1)}(t) F_b^\textrm{(1)}(t) F_c^\textrm{(1)}(t)$. The number $N$ of grid points should be large enough to minimize the noise caused by the discretization. Then spacing $h$ will be sufficiently small to provide a high accuracy of the spatial integration. It is evident that in the limits $L, N \to \infty$ provided $h \to 0$, the discretized equations (17)--(19) coincide with their original, continuous counterparts [Eqs.~(11)--(13)].

As we can see, expressions (17)--(19) are quite cumbersome. Moreover, as was mentioned in the preceding section, we have up 30 independent parameters of the general SMD model even at $M=2$. In order to simplify our further exposition of the algorithm and numerical results obtained by it, we will restrict the discussion in the present paper to a particular one-component ($a,b=M=1$) model in one-dimensional ($d=1$) space with no regular movement, $q_a^\textrm{(M)}=0$, $w_{ab}^\textrm{(M)}=0$, $\mu_a^\textrm{(M)}=0$, and no neighbour-dependent birth, $w_{ab}^\textrm{(B)}=0$. So, the remaining parameters will be related to the intrinsic death $q_a^\textrm{(D)}$, competition $w_{ab}^\textrm{(D)}$ with $c_a^\textrm{(D)}$ and $\sigma_{ab}^\textrm{(D)}$, vegetative birth $q_a^\textrm{(B)}$ as well as dispersal $\mu_a^\textrm{(B)}$ of born agents with $s_{ab}^\textrm{(B)}$. At $M=1$, this reduces the total number of parameters from 11 to 5. As a result, we come to a popular birth-dispersal-death-competition (BDDC) population model which was first introduced in spatial ecology by Bolker and Pacala \cite{BolkerPac, BolkerPaca} as well as by Law, Murrell, and Dieckmann \cite{Law} (aka BDLMP \cite{Birch} or BDLP \cite{FinKon} model). It is known also as the spatial and stochastic logistic model \cite{BolkerPac, Law, FinKonK, OvaskFin, FinKonKoz, OvaskCor} (abbreviated as SLM \cite{FinKonK} or SSLM \cite{OvaskFin}). Our algorithm is quite general in the sense that it can be used for more complex SMD models. The specific BDDC model is considered here only as an illustrative example of efficiency of our numerical approach.

For the BDDC model, the SMD equations (17)--(19) are simplified to
\begin{align}
\frac{{\rm d} F^\textrm{(1)}}{{\rm d} t} =&\ \Big(q^\textrm{(B)} - q^\textrm{(D)} \Big) F^\textrm{(1)} - h \sum_i w_i^\textrm{(D)} \zeta_i F_i^\textrm{(2)} \, , \\[8pt]
\frac{{\rm d} F_i^\textrm{(2)}}{{\rm d} t} =&\ - \ 2 \left( q^\textrm{(D)} + w_i^\textrm{(D)} \right) F_i^\textrm{(2)} + h \sum_j \mu_j^\textrm{(B)} q^\textrm{(B)} \zeta_j \Big( F_{i+j}^\textrm{(2)} + F_{i-j}^\textrm{(2)} \Big) \nonumber \\[1pt] &\ - \ h \sum_j w_j^\textrm{(D)} \zeta_j \Big( F_{ij}^\textrm{(3)} + F_{-i,j}^\textrm{(3)} \Big) + 2 \mu_i^\textrm{(B)} q^\textrm{(B)} F^\textrm{(1)} \, , \\[5pt] \nonumber \frac{{\rm d} F_{ij}^\textrm{(3)}}{{\rm d} t} =&\ - \ \left( 3 q^\textrm{(D)} + 2 w_i^\textrm{(D)} + 2 w_j^\textrm{(D)} + 2 w_{i-j}^\textrm{(D)} \right) F_{ij}^\textrm{(3)} - h \sum_k w_k^\textrm{(D)} \zeta_k \Big( F_{ijk}^{(4)} + F_{-i,j-i,k}^{(4)} + F_{-j,i-j,k}^{(4)} \Big) \nonumber \\[2pt] &\ + \ \Big( \mu_i^\textrm{(B)} \big( F_{j-i}^\textrm{(2)} + F_j^\textrm{(2)} \big) + \mu_{j}^\textrm{(B)} \big( F_i^\textrm{(2)} + F_{j-i}^\textrm{(2)} \big) + \mu_{i-j}^\textrm{(B)} \big( F_i^\textrm{(2)} + F_j^\textrm{(2)} \big) \Big) q^\textrm{(B)} \nonumber \\[9pt] &\ + \ h \sum_k \mu_k^\textrm{(B)} \zeta_k \Big( F_{k+i,k+j}^\textrm{(3)} + F_{k-i,k+j-i}^\textrm{(3)} + F_{k-j,k+i-j}^\textrm{(3)} \Big) q^\textrm{(B)} \, ,
\end{align}
where $i,j=-N,-N+1,\ldots,0,1,2,\ldots,N$, and the component subscripts ($a=b=c=1$) have been omitted. Then the FK closure (20) takes the form
\begin{equation}
F_{ijk}^{(4)} = \frac{F_{ij}^\textrm{(3)} F_{ik}^\textrm{(3)} F_{jk}^\textrm{(3)} F_{j-i,k-i}^\textrm{(3)}}{F_i^\textrm{(2)} F_j^\textrm{(2)} F_k^\textrm{(2)} F_{j-i}^\textrm{(2)} F_{k-i}^\textrm{(2)} F_{k-j}^\textrm{(2)}} \left( F^\textrm{(1)} \right)^4 .
\end{equation}
Eqs.~(21)--(23) constitute a coupled system of autonomous ordinary differential equations with respect to unknown quantities $F^\textrm{(1)}$, $F_i^\textrm{(2)}$, and $F_{ij}^\textrm{(3)}$, where $i,j=-N, \ldots, N$. In general, the number of these equations and quantities is the same and equal to $1+(2N+1)+(2N+1)^2$. Taking into account that it is necessary to perform the summations over $i$, $j$, and $k=-N, \ldots, N$ in the rhs of Eqs.~(21)--(23), the total number of operations for a given time $t$ will be of order of $(2N+1)+(2N+1)^2+(2N+1)^3$, i.e., $\sim 8 N^3$ at $N \gg 1$, that is quite large.

\subsection{Optimization}

Consider now a question of how to reduce the number of operations to a minimum. Fist of all, the number of independent values for the second-order spatial moment decreases from $2N+1$ to $N+1$ because of the symmetry property $F_{-i}^\textrm{(2)}=F_i^\textrm{(2)}$, so that $i=0,1,\ldots,N$ in Eq.~(22). For the same reason, the summation $\sum_{i=-N}^N w_i^\textrm{(D)} \zeta_i F_i^\textrm{(2)}$ in the rhs of Eq.~(21) simplifies to $w_0^\textrm{(D)} \zeta_0 F_0^\textrm{(2)} + 2 \sum_{i=1}^N w_i^\textrm{(D)} \zeta_i F_i^\textrm{(2)}$. Further, it is worth emphasizing one additional symmetry property of the third-order spatial moment, namely, the inversion $F^\textrm{(3)}(-\xi,-\xi',t) = F^\textrm{(3)}(\xi,\xi',t)$. It follows from the fact that the interaction and dispersal kernels are invariant under the substitution $(\xi,\xi') \to (-\xi,-\xi')$. Then $F^\textrm{(3)}_{-i,-j}=F^\textrm{(3)}_{ij}$, so that the sum $\sum_{j=-N}^N w_j^\textrm{(D)} \zeta_j \big( F_{ij}^\textrm{(3)} + F_{-i,j}^\textrm{(3)} \big)$ in the rhs of Eq.~(22) can be replaced by $2 w_0^\textrm{(D)} \zeta_0 F_{i,0}^\textrm{(3)} + 2 \sum_{j=1}^N w_j^\textrm{(D)} \zeta_j \big( F_{ij}^\textrm{(3)} + F_{i,-j}^\textrm{(3)} \big)$. Moreover, $F^\textrm{(3)}_{ij} = F^\textrm{(3)}_{ji}$, so that in view of the inversion $F^\textrm{(3)}_{-i,-j} = F^\textrm{(3)}_{ij}$, the direct calculations in Eq.~(23) should be carried out, in fact, only within the conus $j \ge |i|$ at $i=-N,\ldots,N$. Then the values of $F_{ij}^\textrm{(3)}$ for $i,j$ lying outside this conus can easily be reproduced using the above two symmetry properties, $F_{ji}^\textrm{(3)}=F_{ij}^\textrm{(3)}$ and $F_{-i,-j}^\textrm{(3)} = F_{ij}^\textrm{(3)}$. This will reduce the computation costs almost in four times. Note also that the function $F_{i+k,j+k}^\textrm{(3)}$ is invariant under the replacements $(i,j \to -i,j-i)$ and $(i,j \to -j,i-j)$ at $k=0$. However, this invariance is violated at $k \ne 0$ for $F_{i+k,j+k}^\textrm{(3)}$ and does not take place at all for $F_{ijk}^{(4)}$ at any $k$, including $k=0$. The same invariance is valid for the combination $F_i^\textrm{(2)} F_j^\textrm{(2)} F_{j-i}^\textrm{(2)}$, that can be exploited when handling Eq.~(24). Using all the above tricks, the computational efforts can be reduced nearly on one order in magnitude with the total number of operations of order of $\propto N^3$.

Further important issue is to reduce the number $N$ of grid points as much as possible without loss of precision. At fixed $h$, this can be achieved by decreasing the basic length $L$ because $h=L/N$. The length $L$ can be decreased by additionally reducing the finite-size effects with the help of optimal boundary conditions when mapping the infinite range $\xi \in [0, \infty[$ by the finite area $\xi_i \in [0,L]$. The best way to minimize the finite-size uncertainties consists in choosing asymptotic boundary conditions. Mention that the values $F_i^\textrm{(2)} = F^\textrm{(2)}(\xi_i,t)$ are known only for $i=0,1,\ldots,N$ with $\xi_0=0 < \xi_1 < \ldots < \xi_N=L$. Note also that $F_{-i}^\textrm{(2)} = F^\textrm{(2)}(-\xi_i,t) = F^\textrm{(2)}(\xi_i,t) = F_i^\textrm{(2)}$ owing to the symmetry property. Then, whenever $i$ exceeds the boundary number (i.e., whenever $|i| > N$) during the calculations according to Eqs.~(21)--(24), the values $F_i^\textrm{(2)}$ are replaced by their MF asymptotic counterpart $F_\infty^\textrm{(2)} = \lim_{\xi \to \infty} F^\textrm{(2)}(\xi,t) = F^\textrm{(1)}(t) F^\textrm{(1)}(t)$. The situation with the third-order distribution function $F^\textrm{(3)}(\xi,\xi',t)$ is somewhat more complicated. First of all, the values $F_{ij}^\textrm{(3)} = F^\textrm{(3)}(\xi_i,\xi_j,t)$, where $i,j=-N, -N+1, \ldots, 0, 1, \ldots, N-1, N$, satisfy up six symmetry properties, $F_{ij}^\textrm{(3)} = F_{ji}^\textrm{(3)} = F_{-i,j-i}^\textrm{(3)} = F_{j-i,-i}^\textrm{(3)} = F_{-j,i-j}^\textrm{(3)} = F_{i-j,-j}^\textrm{(3)}$, plus the inversion $F_{-i,-j}^\textrm{(3)} = F_{ij}^\textrm{(3)}$. Then, whenever the both numbers $|i|$ and $|j|$ or one of them exceed $N$ during the calculations, the values of $F_{ij}^\textrm{(3)}$ can be obtained in the following way. First, we try to reproduce them explicitly using the symmetry properties. This is indeed possible for most pairs $(i,j)$, because the numbers $|i-j|$ can appear to be within the explicit interval, while $i$ and $j$ separately or together to be outside of it. For the rest pairs we apply the KSA boundary condition $F_{ij}^\textrm{(3)}=F_i^\textrm{(2)} F_j^\textrm{(2)} F_{|i-j|}^\textrm{(2)}/(F^\textrm{(1)} F^\textrm{(1)} F^\textrm{(1)})$, where $F_i^\textrm{(2)}$ and/or $F_j^\textrm{(2)}$ are the explicit values of the second-order function or their asymptotic counterparts if $|i|$ or/and $|j|$ exceed $N$. Alternatively, when at least one of the number $|i|$ or $|j|$ exceed $N$, the values $F_{ij}^\textrm{(3)}$ can be replaced by their MF asymptotic $\lim_{|\xi|, |\xi'| \to \infty} F^\textrm{(3)}(\xi,\xi',t)=F^\textrm{(1)}(t) F^\textrm{(1)}(t) F^\textrm{(1)}(t)$. Investigations show that the method with the KSA boundary condition is superior with respect to the MF asymptotic. Such a superiority means that then with increasing $N$ a faster convergence of the investigated quantities to their exact values (corresponding to $N \to \infty$) can be provided.

Additional way to decrease $N$ at fixed $L$ is to apply large values of $h=L/N$. Such values indeed can be employed without loss of precision by choosing the best method for numerical spatial integration. In particular, instead of the simple composition trapezoidal scheme with $\zeta_0 = \zeta_1 = \zeta_2 = \ldots = \zeta_{N-1} = 1$ and $\xi_N=1/2$, we can use the more accurate composition Simpson rule, where $\zeta_N=1/3$, $\zeta_{N-1}=4/3$, $\zeta_{N-2}=2/3$, $\zeta_{N-3}=4/3$, $\zeta_{N-4}=2/3$, \ldots, $\zeta_0=4/3$ if $N$ is odd and $\zeta_0=2/3$ if $N$ is even. Note that the weights satisfy the normalization condition $\sum_{i=-N}^N \zeta_i = \zeta_0 + 2 \sum_{i=1}^N \zeta_i= 2 N$. More complicated schemes for numerical integration in coordinate space (which take into account explicitly the Gaussian form of the dispersal and competition kernels) can also be involved.

Moreover, it is necessary to take into account that the dispersal $\mu^\textrm{(B)}(\xi)$ and competition $w^\textrm{(D)}(\xi)$ kernels tend to zero with increasing the separation between agents, so that at $|\xi| > R_\textrm{D,B}$ their influence can be neglected. The truncation radiuses can be cast in the forms $R_\textrm{D}=Q \sigma^\textrm{(D)}$ and $R_\textrm{B}=Q s^\textrm{(B)}$, where $Q \gg 1$. In the case of the Gaussian kernels (15) we can take $Q=5-6$ because then $\exp(-Q^2/2) \sim 4 \cdot 10^{-6} - 1.5 \cdot 10^{-8} \approx 0$. For compensation, we can use a slight correction of the kernels to the form $\tilde \mu^\textrm{(B)}(\xi) = \mu^\textrm{(B)}(\xi)/Q_\mu$ and $\tilde w_{ab}^\textrm{(P)}(\xi) = w_{ab}^\textrm{(P)}(\xi)/Q_w$, where $Q_\mu=\int_{-R_\textrm{B}}^{R_\textrm{B}} \mu^\textrm{(B)}(\xi) {\rm d} \xi \approx 1$ and $Q_w=\int_{-R_\textrm{D}}^{R_\textrm{D}} w^\textrm{(D)}(\xi) {\rm d} \xi \approx 1$. Then the normalization conditions for the corrected kernels will be satisfied exactly despite the truncation of the latter. For the top-hat kernels (16) we have $Q=1$ since they are truncated by definition, so that no corrections are then required. In view of the truncation, the upper integer $N$ can be replaced by its smaller values $N_\textrm{D,B} = R_\textrm{D,B}/h < N$ when performing the summations with $w_i^\textrm{(D)}$ or $\mu_i^\textrm{(B)}$ at $i \le N_\textrm{D,B}$ in Eqs.~(21)--(23) provided $R_\textrm{D,B} < L$. As a consequence, further decrease of the computational expenses can be expected. Note also that the kernel values $w_i^\textrm{(D)}$ and $\mu_i^\textrm{(B)}$ can be calculated once in advance on the very beginning before time integration, because the grid points $\xi_i$ remain unchanged in time. This will also speed up the compuations.

\subsection{Time integration}

Finally, let us introduce a scheme for the time integration of the coupled system of autonomous ordinary differential equations (21)--(23). We first cast them in the following compact form
\begin{equation}
\frac{{\rm d} \Gamma}{{\rm d} t} = \dot \Gamma = \Phi(\Gamma) \, ,
\end{equation}
where $\Gamma(t)=\{ F^\textrm{(1)}, F_i^\textrm{(2)}, F_{ij}^\textrm{(3)} \}$ is a space of dynamical variables and $\Phi(\Gamma)$ is a function of them. The explicit form of $\Phi(\Gamma)$ is defined by the set of expressions in the rhs of Eqs.~(21)--(23).

Many approaches were derived to perform the time integration of differential equations. They include explicit and implicit integrators of various precisional orders in time step. Among them it is worth mentioning the classical explicit Runge-Kutta scheme of the fourth order (RK4). Although the latter is not capable \cite{OmelKoz} for spatially inhomogeneous birth-death systems, it is quite good in the homogeneous limit. RK4 is commonly exploited in spatial population dynamics of lattice \cite{Baker, Markham, Markhama, Simpsona, Johnston} and off-lattice \cite{Middleton, Matsiaka} models. Higher-order schemes can also be involved, but they will require more computational efforts.

The RK4 scheme reads
\begin{equation}
\Gamma(t+\Delta t) = \Gamma(t) + \Big( \Phi_1 + 2 \Phi_2 + 2 \Phi_3 + \Phi_4 \Big) \frac{\Delta t}{6} + \mathcal{O}(\Delta t^5) \, ,
\end{equation}
where
\begin{equation}
\Phi_1 = \Phi\big(\Gamma(t)\big) \, , \ \ \ \Phi_2 = \Phi\big(\Gamma(t)+\Phi_1 \Delta t/2\big) \, , \ \ \ \Phi_3 = \Phi\big(\Gamma(t)+\Phi_2 \Delta t/2\big) \, , \ \ \ \Phi_4 = \Phi\big(\Gamma(t)+\Phi_3 \Delta t\big)
\end{equation}
relate to four intermediate states and $\Delta t$ denotes the time step. Alternatively, to reduce the computational efforts, we can use the second-order RK scheme (midpoint rule)
\begin{equation}
\Gamma(t+\Delta t) = \Gamma(t) + \Phi\big(\Gamma(t)+\Phi_1 \Delta t/2\big) \Delta t + \mathcal{O}(\Delta t^3) \, .
\end{equation}

Starting from an initial configuration $\Gamma(0)$ and repeatedly applying Eq.~(26) or (28) the corresponding number of times, the numerical solution $\Gamma(t)$ can be found for any $t >0$. Of course, Eq.~(26) or (28) are not exact, so that the $\mathcal{O}(\Delta t^5)$- or $\mathcal{O}(\Delta t^3)$-uncertainties arise. However, they can be reduced to arbitrary small values by decreasing the length $\Delta t$ of the time step.

\section{Results and Discussion}

\subsection{Computational details}

The SMD/FK calculations were carried for the BDDC model using two characteristic sets I and II of parameters. In the first one consisting of four subsets, we have the same values for the intrinsic birth $q^\textrm{(B)} = 0.6$ and death $q^\textrm{(D)} = 0.1$ rates as well as for the competition strength $c^\textrm{(D)} = 0.01$, but four different combinations for the radiuses of the competition and dispersal Gaussian kernels, namely, (a) $\sigma^\textrm{(D)} = 0.02$, $s^\textrm{(B)} = 0.12$; (b) $\sigma^\textrm{(D)} = 0.12$, $s^\textrm{(B)} = 0.12$; (c) $\sigma^\textrm{(D)} = 0.02$, $s^\textrm{(B)} = 0.02$; and (d) $\sigma^\textrm{(D)} = 0.12$, $s^\textrm{(B)} = 0.02$. These four situations are very similar to those described in Ref.~\cite{Law} when performing IBM simulations for the BDDC model. In the second set consisting of two subsets we have that $q^\textrm{(B)} = 1$, $q^\textrm{(D)} = 0.01$, and $c^\textrm{(D)} = 1$ with two different radiuses combinations: (a) $\sigma^\textrm{(D)} = 0.1$, $s^\textrm{(B)} = 1$ for the Gaussians; and (b) $\sigma^\textrm{(D)} = 1$, $s^\textrm{(B)} = 0.1$ for the top-hat kernels. The latter two subsets coincide completely with those used recently by us \cite{OmelKoz} when considering the BDDC model for inhomogeneous systems within the KSA closure. Such two subsets should be considered as a more aggressive choice for stress testing of the KSA and FK approaches because they lead to much stronger disaggregated and aggregated spatial structures, respectively, than in the case of the first set.

The SMD/FK equations (21)--(23) were solved using our numerical algorithm derived in the preceding Section 3. The basic length and number of grid points for the first set were equal correspondingly to $L=2$ and $N=500$, except the second subset where $N=100$. In the case of the two subsets of the second set they were equal to $L=12$ with $N=600$ and $L=4$ with $N=400$, respectively. Spatial integration was performed with help of the composition Simpson rule. The Gaussian kernels were truncated at a reduced length of $Q=6$. Time integration was done employing the RK4 scheme with a step of $\Delta t=0.05$ in all the cases. Further increasing space and time resolution does not affect the solutions. Even smaller values of $L$ and $N$ could be chosen without loss of precision.

For the first set, the initial ($t=0$) density was put to be $F^{\rm (1)}(0) = 16\exp[0] / (2\pi)^{1/2} \approx 6.38$ that is equal to the amplitude of the initial spatially inhomogeneous Gaussian distribution $F^{\rm (1)}(x,0) = \nu_0 \exp[-x^2/(2 \sigma_0)] / (2\pi)^{1/2}$ [cf. Eq.~(15)] at $x=0$ with $\sigma_0 = 1$ and $\nu_0=16$ (which will be used in our future IBM simulations of spatially inhomogeneous systems with $\nu_0$ entities at $t=0$). Now, when considering the spatially homogeneous case, the above initial value was chosen for the sake of convenience (when comparing the results with inhomogeneous data in our next papers). For the second set, the initial density was $F^{\rm (1)}(0)=\exp[0]/(2\pi)^{1/2} \approx 0.4$.

For the purpose of validation of our SMD/FK approach, we have also carried out IBM simulations using the direct method of the stochastic simulation algorithm (SSA) by Gillespie \cite{Gillespie, Gillespei}, aka the dynamical Monte-Carlo method. In the case of the first set, the simulations were started at $t=0$ from $\mathcal{N}_0=64$ entities distributed at random for each ensemble (realization). The (fixed) length of the simulation box was chosen to be $\mathcal{L}=\mathcal{N}_0/F^{\rm (1)}(0) \approx 10$ to provide the same initial density $F^{\rm (1)}(0)$ as in the case of the SMD/FK calculations. The total number of time steps (which are random in length and decrease with increasing the size of the system) and the total number $\mathcal{K}$ of ensembles were equal to 20 000/20 000, 32 000/20 000, 180 000/10 000 and 180 000/10 000 for the first, second, third, and fourth subsets, respectively. Such large values of $\mathcal{K}$ are needed to reduce the statistical noise. For the second set, we have chosen $\mathcal{N}_0=16$ with $\mathcal{L} \approx 40$ and $\mathcal{N}_0=64$ with $\mathcal{L} \approx 16$ to provide the initial densities $F^{\rm (1)}(0) = \exp[0]/(2\pi)^{1/2} \approx 0.4$ and $F^{\rm (1)}(0) = 10 \exp[0]/(2\pi)^{1/2} \approx 4$, for the first and second subsets, respectively. In the last two cases, the total number of time steps and ensembles were equal correspondingly to 12 000/200 000 and 20 000/1 000 000. The toroidal boundary conditions were used to minimize the finite-size effects. Further increasing $\mathcal{L}$ and $\mathcal{K}$ does not effect the results. For this reason, they can be treated as ``exact'' (to within a statistical noise which was estimated in our IBM simulations to be of order of 0.0002, see below) or etalon for comparison with theoretical approaches.

In the SMD/FK calculations it is necessary also to determine initial values for the second- and third-order moments. Since in the SSA simulations the entities are distributed at random on the very beginning, the obvious choice is the uncorrelated initial conditions $F^\textrm{(2)}(\xi,0) = \zeta^\textrm{(2)} F^\textrm{(1)}(0) F^\textrm{(1)}(0)$ and $F^\textrm{(3)}(\xi,\xi',0) = \zeta^\textrm{(3)} F^\textrm{(1)}(0) F^\textrm{(1)}(0) F^\textrm{(1)}(0)$. At $\zeta^\textrm{(2)}=1$ and $\zeta^\textrm{(3)}=1$ this corresponds to the asymptotic (MF) approximation of the second- and third-order correlations for the infinite system. Then the SMD/FK and SSA time dependencies of the spatial moments will be synchronized between themselves. In our case, the system in the IBM simulations is finite with the initial number $\mathcal{N}_0$ of entities. The finite-size corrections read $\zeta^\textrm{(2)} \approx (\mathcal{N}_0 - 1) / \mathcal{N}_0$ if $\mathcal{N}_0 \ge 2$ and $\zeta^\textrm{(3)} \approx (\mathcal{N}_0 - 1) (\mathcal{N}_0 - 2) / \mathcal{N}_0^2$ if $\mathcal{N}_0 \ge 3$. Otherwise, they are equal to zero, meaning that no pair or not triplet can be formed with less than 2 or 3 particles, respectively. Strictly speaking, the choice of initial values for $\Gamma(0)$ is not so important when computing the spatial moments in the steady state. Indeed, even putting $\zeta^\textrm{(2)} = \zeta^\textrm{(3)} = 0$ with zeroth initial second- and third-order correlations, the latter are quickly reproduced owing to the interactions and we come to the same state for any $\Gamma(0) > 0$. This choice will only (slightly) influences on the relaxation time $\tau$ needed to achieve this state from an initial configuration. In our IBM dynamical Monte-Carlo simulations with $\mathcal{N}_0=16$ or 64 we have that $\zeta_2 \approx 0.94$ or 0.98 and $\zeta_3 \approx 0.82$ or 0.95 on the very beginning. At the end ($t \approx 2 \tau$) the mean number of entities increased significantly, nearly to $\mathcal{N}_t \sim 430-560$ for the fist set and 120 for the first subset of the second set, resulting in $\zeta_2 = \zeta_3 \approx 1$ (the second subset of the second set, where $\lim_{t \to \infty} \mathcal{N}_t \to 0$, presents a special case, see subsection {\em 4.4}). This means that no finite-size corrections of the second and third spatial moments are necessary in the steady state during the IBM simulations. No such corrections are alse needed for the SMD/FK calculations because the finite-size effects were already taken into account by the derived algorithm (see subsection {\em 3.2}).

In the SSA simulations, the mean value of any single observable quantity was obtained by averaging it values over all statistical ensembles ($\kappa=1,2,\ldots,\mathcal{K}$). The binary correlation function $g^\textrm{($\kappa$)}(\xi,t)$ in the $\kappa$-th ensemble at given $\xi$ and $t$ was calculated by counting the number $\delta N^\textrm{($\kappa$)}_\iota(\xi,t)$ of pairs around each ($\iota = 1, 2, \ldots, \mathcal{N}^\textrm{($\kappa$)}$) entity with the interparticle separation lying in the narrow interval $[\xi-\delta \xi/2, \xi+\delta \xi/2]$. During this counting, the state of the system should correspond to the narrow time interval $[t-\delta t/2, t+\delta t/2]$. This leads to the coarse-grained averaging $g^\textrm{($\kappa$)}(\xi,t) = \frac{1}{\delta \xi \delta t \mathcal{N}^\textrm{($\kappa$)}} \sum_\iota \delta N^\textrm{($\kappa$)}_\iota(\xi,t)$. A spacing of $\delta \xi= 0.002 - 0.005 \ll \min({s^\textrm{(B)}, \sigma^\textrm{(B)}})$ and a step of $\delta t=0.1 \ll \tau$ were chosen when performing the coarse graining in space and time, respectively. Analogous coarse-grained averaging was performed for all other time or/and space-dependent quantities. Expected values for them were then obtained by carrying out statistical averaging over the realizations. In particular, for the pair correlation function one finds that $g(\xi,t) = \sum_\kappa g^\textrm{($\kappa$)}(\xi,t)/\mathcal{K}$. Additional averaging over time was performed in the steady state $t > \tau$ (where $g(\xi,t) \equiv g(\xi)$ and all other observables become time independent) to decrease the statistical noise.

For the purpose of comparison, the SMD calculations with the power-3 KSA closure [see Eq.~(B2)] were also performed. During these calculations only two (not three) equations (21) and (22) were solved, while the third-order moment was approximated using the discretized KSA form $F_{ij}^\textrm{(3)} = F_i^\textrm{(2)} F_j^\textrm{(2)} F_{j-i}^\textrm{(2)}/(F^\textrm{(1)} F^\textrm{(1)} F^\textrm{(1)})$. The power-2 closure, which expresses the third moment in terms of weighted sums of lower-order moments [see Eq.~(B1)] was considered, too. Its discretized form reads $F_{ij}^\textrm{(3)} = [ \alpha F_i^\textrm{(2)} F_j^\textrm{(2)}/F^\textrm{(1)} + \beta F_i^\textrm{(2)} F_{j-i}^\textrm{(2)}/F^\textrm{(1)} + \gamma F_j^\textrm{(2)} F_{j-i}^\textrm{(2)}/F^\textrm{(1)} - \beta F^\textrm{(1)} F^\textrm{(1)} F^\textrm{(1)} ] / (\alpha+\gamma)$, where $\alpha=\beta=\gamma=1$ for the symmetric version and $\alpha=4$, $\beta=\gamma=1$ for the asymmetric one \cite{Law}. Finally, in the MF approximation it is necessary to solve exclusively one equation (21) complemented by the simplest (Poisson) closure $F_{ij}^\textrm{(2)}=F^\textrm{(1)} F^\textrm{(1)}$ for the second-order spatial moment.

\subsection{First spatial moment (set {\rm I})}

The first-order spatial moment $n(t) \equiv F^\textrm{(1)}(t)$ of the BDDC model is plotted in Fig.~1 as a function of time for four subsets (a), (b), (c), and (d) of the parameters taken from the first set I. Remember that $n(t)$ is the spatial mean density of entities in the population at time $t$. The SMD results obtained using the power-4 FK, power-3 KSA, symmetric power-2 (SPW), asymmetric power-2 (APW), and MF closures are shown as bold black, green, magenta, thin black, and cyan curves, respectively. The IBM simulations data are presented by the blue and red curves. The first one corresponds to instantaneous stochastic values $n^\textrm{($\kappa$)}(t)$ related to a single, randomly chosen $\kappa$-realization. The red curve refers to values $n(t)=\frac{1}{\mathcal{K}} \sum_{\kappa=1}^\mathcal{K} n^\textrm{($\kappa$)}(t)$ averaged over large numbers $\mathcal{K} \sim 10 000 - 1 000 000$ of statistical ensembles. Since the system is finite, we can observe visible density fluctuations in $n^\textrm{($\kappa$)}(t)$ which behave randomly according to the stochastic nature of the BDDC model in IBM simulations. It should be emphasized that the spatial moments describe properties of the system in terms of the ensemble average of stochastic agent-based (microscopic) behaviour. These moments (yielding expected values for distribution functions) do not give information on the size or nature of fluctuations around that ensemble average, and they cannot, for instance, be used to estimate the probability that a population will eventually go extinct. That is why the spatial moments satisfy the (mesoscopic) deterministic master equations without any explicit stochastic contributions.

\begin{figure}[t]
\centering
\includegraphics[width=0.56\textwidth]{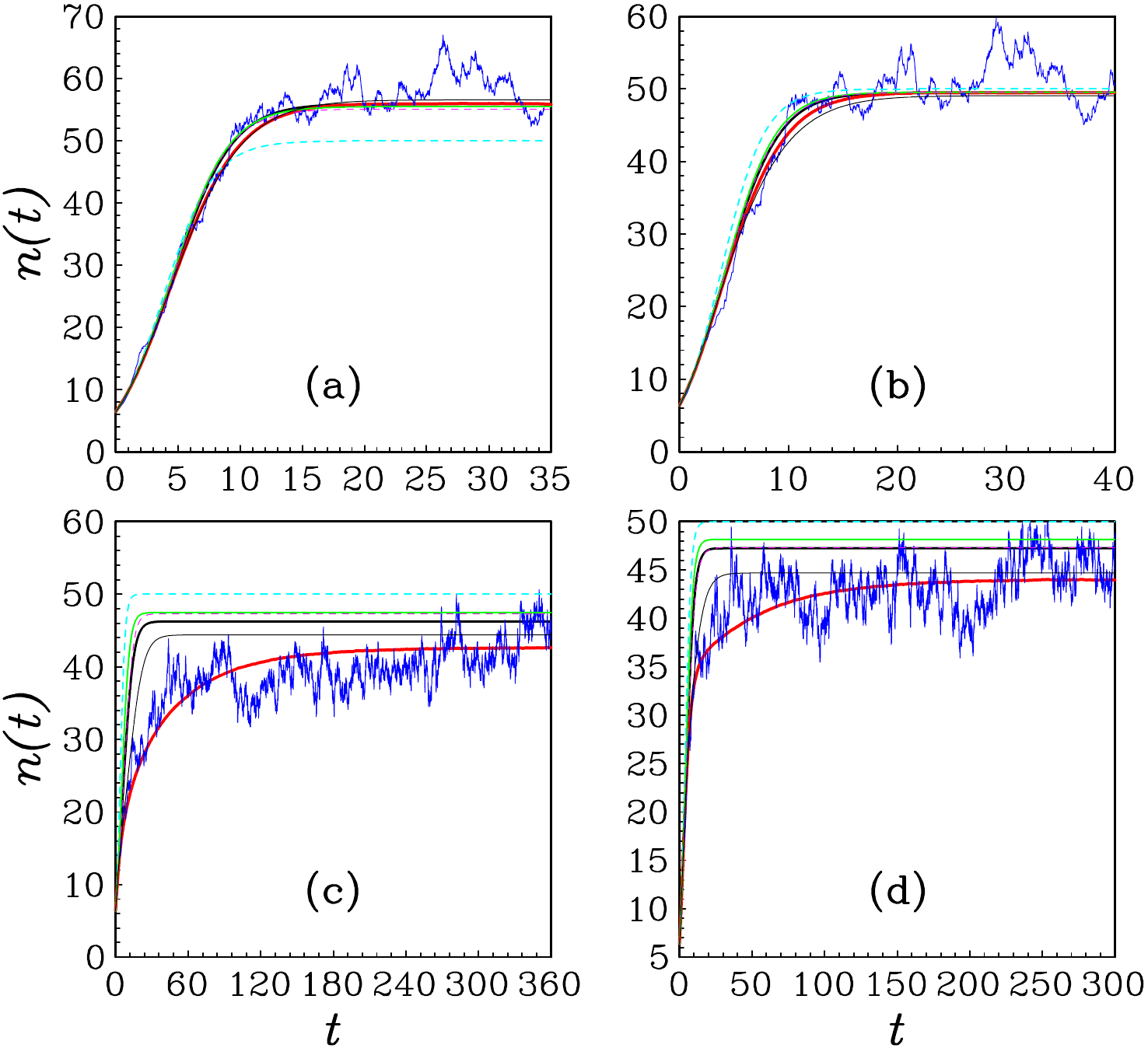}
\caption{First spatial moment (mean density) as a function of time obtained by the SMD approach for the BDDC model using the FK (bold black), KSA (green), SPW (magenta), APW (thin black), and MF (cyan) closures in comparison with that of the IBM simulations (red and blue). Parts (a), (b), (c), and (d) correspond to four different subsets of the model parameters from the first set (see the text).}
\label{f1}
\end{figure}

All the curves in Fig.~1, starting at $t=0$ from the same initial value $n(0) \approx 6.38$, begin to increase with different rapidity for each subset. After a relaxation time $\tau$, the function $n(t)$ soon achieves a steady-state regime in which $d n/d t=0$ with $n(t) = n_{\rm s}$ at $t > \tau$. The values of $\tau$ and the density $n_{\rm s}$ in the steady sate depend on the choice of the parameters of the model and on the approximation used for the description. According to the ``exact'' IBM simulation results, we have $n_{\rm s} \sim 42-56$ with $\tau \approx 20$ for the cases (a) and (b), while $\tau \approx 200$ for the situations (c) and (d). In the steady-state regime we have dynamical equilibrium when the number of born agents is equal in average to the number of died agents. If the initial density is much smaller than its steady-state value (i.e., $n(0) \ll n_{\rm s}$, as in our case), the population begins to quickly grow because a small mortality caused by the competition (the latter is proportional to the density). In the opposite limit $n(0) \gg n_{\rm s}$ of very dense initial configurations, the competition mortality will dominate over the born processes and, as a consequence, the number of agents will rapidly decrease on the very beginning. In both cases, the system will come to the same steady-sate value $n(t) \approx n_{\rm s}$ on sufficiently long times $t > \tau$, where the relaxation time $\tau$ depends on the initial condition $n(0)$. The trivial zeroth solution $n(t)=0$ is not stable, so that even at very small positive values $n(0) \approx 0$, the system sooner or later will always achieve the steady state. More detailed explanation of the time dependence $n(t)$ in terms of the model parameters was already given earlier (see, e.g., \cite{Law}). In the present study our discussion will be focused mainly on the comparison of different closure schemes in context of their ability for a quantitative description within the SMD approach.

As can be seen from Fig.~1, the MF approximation yields the worst results for $n(t)$. It can be used only in the case when the range of dispersal is nearly equal to that of competition (i.e., when $s^\textrm{(B)} \sim \sigma^\textrm{(D)}$) and the both are not too small [the situation shown in part (b) of the figure when $\sigma^\textrm{(D)} = s^\textrm{(B)} = 0.12$]. Then the departures from spatial Poisson processes can be neglected, resulting in no spatial structure. Besides this special case, the MF theory is too inaccurate. Other approaches are more accurate and their precision increases with increasing the power of the closure. For instance, the FK predictions are always better than those of KSA and SPW. In particular, the FK and IBM curves are practically indistinguishable for segregated [see part (a) where $s^\textrm{(B)} \gg \sigma^\textrm{(D)}$] and weakly aggregated [part (b) with large $s^\textrm{(B)} \sim \sigma^\textrm{(D)}$] states. For moderate and strong aggregations [see parts (c) with small $s^\textrm{(B)} \sim \sigma^\textrm{(D)}$ and (d) when $s^\textrm{(B)} \ll \sigma^\textrm{(D)}$] the deviations between the APW and IBM data are somewhat smaller than those between the FK and IBM ones. However, the APW closure was artificially constructed [see Eq.~(B1)] involving up three adjustable parameters ($\alpha, \beta, \gamma$) to give a closer fit to the average of stochastic realizations for populations that develop strong spatial aggregation \cite{Law}. Moreover, this closure destroys symmetrical properties of the triplet correlation function and can lead, in general, to unphysical negative values for the density of the system (see Appendix B). Being good in the specific region of parameters due to a fortunate cancellation of errors, it performs worse, in general, especially in segregated spatial structures (see next subsections where the results on the second spatial moment are discussed).

From parts (a) and (b) of Fig.~1 we see also that in segregated and weakly aggregated states the SMD/FK approach is able to describe almost quantitatively not only the steady state (dynamical equilibrium) value $n_{\rm s} = n(t)$ of entity density at $t > \tau$ but the time dependence $n(t)$ in the non-equilibrium regime $t \le \tau$. Indeed, the FK and IBM curves are very close to each other. For the moderately and strongly aggregated states [parts (c) and (d) of Fig.~1] none of the closures allow quantitative description of the time dependence. Here we can talk exclusively about a qualitative reproduction because the deviations with respect to the IBM functions are too large, especially at intermediate times $t < \tau/2$. The deviations decrease, however, when approaching the steady-state regime ($t > \tau$) on long times.

\subsection{Second spatial moment (set {\rm I})}

The normalized second-order spatial moment $g(\xi,t) = F^\textrm{(2)}(\xi,t) / (F^\textrm{(1)}(t))^2$ (aka pair or binary correlation function) is shown in Fig.~2 in the steady state regime $g(\xi) = g(\xi,t)$ at $t > \tau$ for moderately segregated (a), weakly aggregated (b) mildly aggregated (c) and strongly aggregated (d) spatial patterns which correspond to the subsets (a), (b), (c) and (d) of the parameters from the first set I. The disaggregation and aggregation (clustering) can be explained by an interplay between the dispersal and competition interactions in the presence of the birth and death processes. Indeed, at $\sigma^\textrm{(D)} \ll s^\textrm{(B)}$ the competition interactions acting over the narrow interval $\xi < \sigma^\textrm{(D)}$ are local and strong. As a result, a sizeable part of agents in this interval dies immediately after their birth, $0 < g(\xi) < 1$, while survivors are overdispersed up to long distances $\xi \gtrsim s^\textrm{(B)}$ with moderate pair correlations $g(\xi) > 1$. This picture is seen in part (a) of Fig.~2. In the opposite regime, when the distance over which offspring disperse is made shorter (by reducing $s^\textrm{(B)}$), agents are increasingly clustered in space, $g(\xi) \gg 1$, around points where they were born. That portion of entities which has dispersed outside the narrow interval $\xi<s^\textrm{(B)}$ is soon killed, $g(\xi) < 1$, by the neighbouring agents owing to the competition with them in the wide domain $s^\textrm{(B)} < \xi < \sigma^\textrm{(D)}$, leading to a disaggregation. Such a pattern is observed in part (d) of Fig.~2, where $\sigma^\textrm{(D)} \gg s^\textrm{(B)}$.

\begin{figure}[t]
\centering
\includegraphics[width=0.56\textwidth]{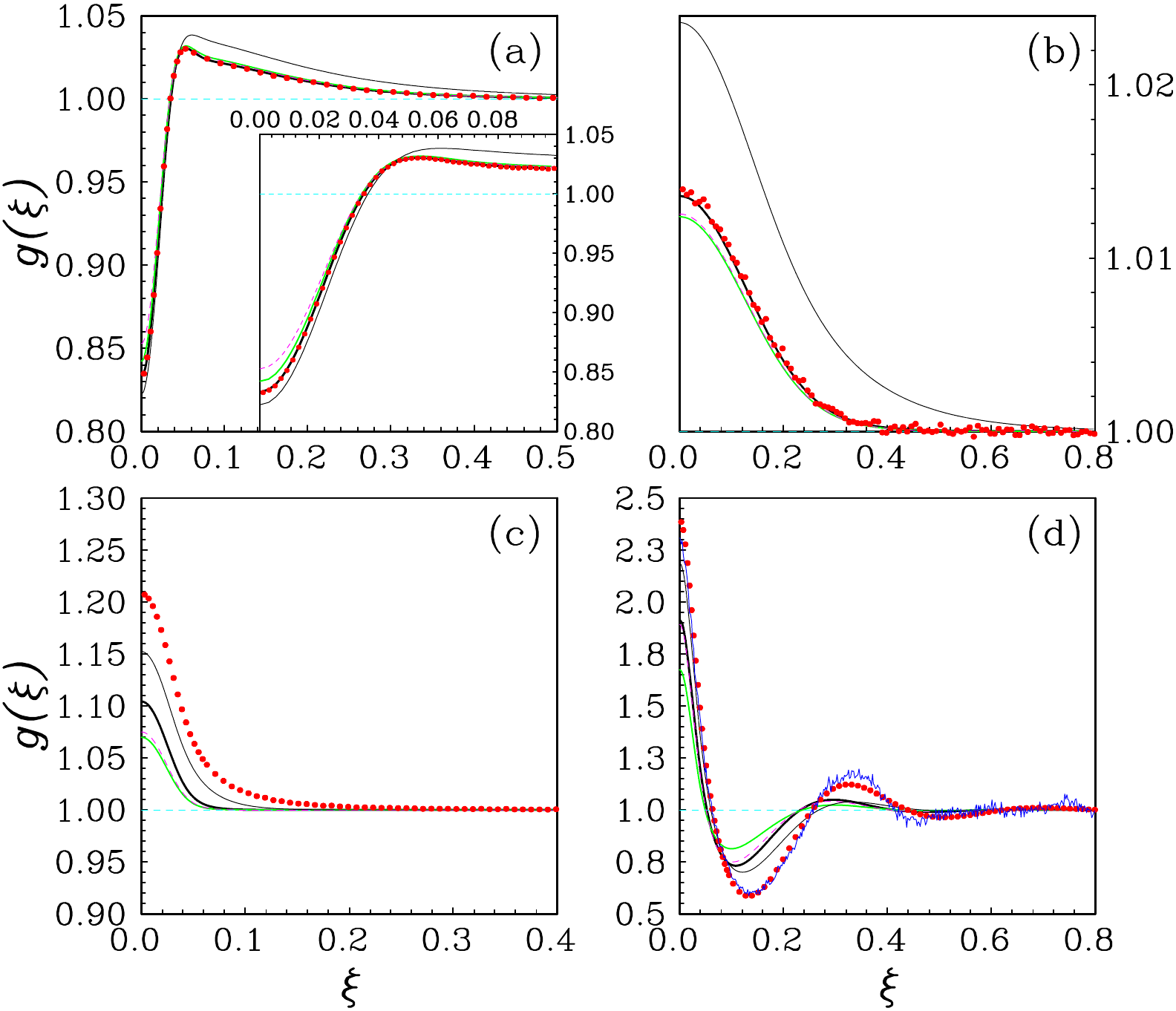}
\caption{Pair correlation function as depending on distance between entities obtained by the SMD approach in the steady state for the BDDC model using the FK (bold black), KSA (green), SPW (magenta), APW (thin black), and MF (cyan) closures in comparison with that of the IBM simulations (red circles). Parts (a), (b), (c), and (d) correspond to four different subsets of the model parameters from the first set (see the text).}
\label{f2}
\end{figure}

In view of Fig.~2 we can state again (as in the case of Fig.~1) that the FK approach is ideal for the description of segregated and weakly aggregated states. Indeed, as can be seen from parts (a) and (b) of Fig.~2, the FK-curves for $g(\xi)$ coincide completely with those of the IBM simulations. At the same time, all other approaches perform evidently worse, especially the APW closure with its largest uncertainties at intermediate and long distances $\xi$. For moderate [part (c)] and strong [part (d)] clustering the accuracy of the FK approach decreases but remains to be higher than that of KSA and SPW. Only APW curves are somewhat closer to those of IBM, but as was said above, this was achieved by an unphysical adjustment of the weighted coefficients in the APW closure for the concrete case of strong aggregation. Note also that the MF theory completely neglects the pair and higher-order spatial correlations, meaning that $g(\xi) \equiv 1$ (see dashed horizontal lines). Fig.~2 demonstrates that $g(\xi)$ can deviate significantly from the MF value 1. At sufficiently long distances $\xi \gtrsim 0.3-0.8$, the pair correlation function tends to its asymptotics $g(\xi) \to 1$ which coincides with the MF value. Remember that $g(\xi)$ is the probability of finding a pair of entities with distance $\xi$ between them relative to the probability of having entities at the same distance if they were independently distributed. Moreover, it satisfies the integral relation $\int_0^L g(\xi,t) {\rm d} \xi / L = \langle \mathcal{N} (\mathcal{N}-1) \rangle / \langle \mathcal{N} \rangle^2$ denoting the ratio of the mean number of pairs to the square of the mean number of entities in the system. This relation follows from the connection $\int F^\textrm{(2)}(\xi,t) {\rm d} \xi = (\langle \mathcal{N}^2 \rangle / \langle \mathcal{N} \rangle - 1) F^\textrm{(1)}(t)$ between the second- and first-order spatial moments (see Appendix C). In the limit of the infinite system $L \to \infty$, $\mathcal{N} \to \infty$ provided $\mathcal{N}/L \to \textrm{const}$ the above ratio tends to 1, confirming the MF asymptotics $g(\xi) \to 1$ at $\xi \to \infty$.

It should be pointed out that the IBM functions $g(\xi)$ plotted in parts (a), (c), and (d) of Fig.~2 are smooth enough since they were averaged over large numbers of ensembles. If we retrieve these functions from a single realization, the statistical noise will become visible. As an example, we included the corresponding result in subset (d) of the figure (see the blue curve). But even for single realizations this noise is relatively small because of large numbers ($\mathcal{N} \sim 500$) of entities in our IBM simulations in the steady states. It will increase for systems with smaller number of entities due to the increased density fluctuations. Only at the scale of Fig.~2(b), the reduced statistical uncertainties (after averaging over the ensembles) are visually observable and can be estimated to be equal of order of 0.0002. Approximately with the same precision all other quantities were calculated in the IBM simulations for all the parameters of the BDDC model considered.

For completeness of our investigation we present also the time-dependent pair correlation functions $g(\xi,t)$ obtained by the SMD/FK approach in the non-equilibrium regime $t \le \tau$ for subsets (a) and (b) of the parameters from the first set. The results on this are shown in parts (a) and (b) of Fig.~3, respectively. The IBM simulations data are also included there. As can be seen, starting from the initial distribution $g(\xi,0)=1$, the non-equilibrium function $g(\xi,t)$ suddenly changes its form at $t=3$ with huge deviations from 1. During the next time intervals $t=6$, 9 and 12 we can watch a convergence of the coordinate dependence of $g(\xi,t)$ to the steady-state behaviour $g(\xi)$ at $t>\tau=20$, so that already at $t=12$ the both curves are practically indistinguishable between themselves. A similar time convergence is observed for the IBM functions.

\begin{figure}
\centering
\includegraphics[width=0.56\textwidth]{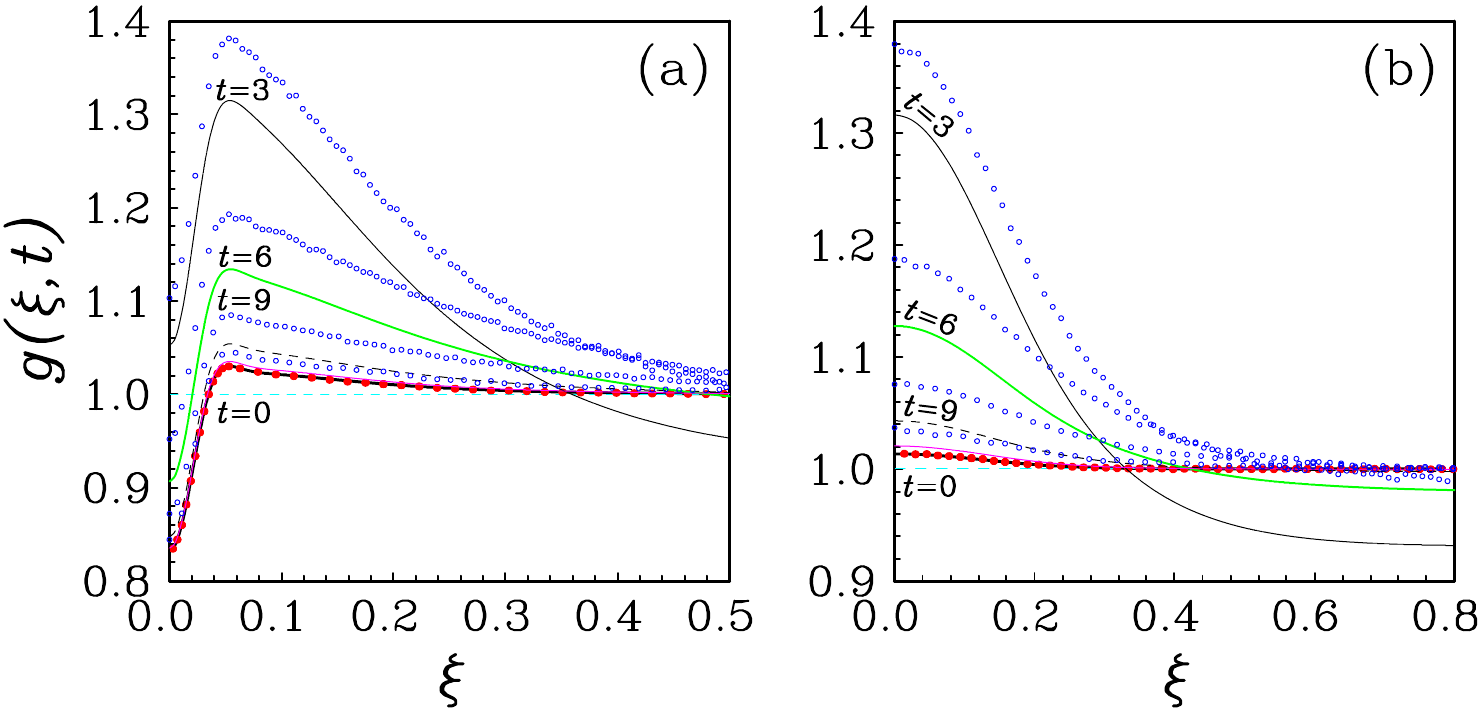}
\caption{Pair correlation function as depending on distance obtained by the SMD/FK approach in the non-equilibrium regime for the BDDC model at five different times $t=0$ (horizontal dashed), $t=3$ (think black), $t=6$ (green), $t=9$ (dashed) and $t=12$ (magenta) versus the steady-state dependence ($t>20$, bold black curves). The IBM simulations data are shown as open and filled circles. Parts (a) and (b) correspond to the first two subsets of the model parameters from the first set.}
\label{f3}
\end{figure}

\subsection{First and second spatial moments (set {\rm II})}

Consider now the two subsets (a) and (b) from the second set II of the model parameters. They correspond to strongly segregated and deeply aggregated states, respectively. Mention that in the first set, the competition-dispersal range ratio $\sigma^\textrm{(D)} / s^\textrm{(B)}$ was varied from 1/6 through 1 to 6. For the second set we moderately increase the intrinsic birth $q^\textrm{(B)}$ from 0.6 to 1.0, significantly raise the intrinsic death $q^\textrm{(D)}$ from 0.1 to 1 and substantially increase the competition strength $c^\textrm{(D)}$ from 0.01 to 1. In addition, the competition-dispersal range ratio $\sigma^\textrm{(D)} / s^\textrm{(B)}$ will be equal to 1/10 (strongly segregated) and 10 (deeply aggregated). Moreover, in the latter case we used the discontinuous top-hat kernels (not the continuous Gaussians). The aim of all these changes is to test the different closures in such an extremely stress situation for detecting their possible limitations.

The first-order spatial moment $n(t)$ obtained by the SMD approach is plotted in Fig.~4 as a function of time. The SMD pair correlation function $g(\xi)$ representing the normalized second-order spatial moment is shown in Fig.~5 in the steady state regime $g(\xi) = g(\xi,t)$ at $t > \tau$. Parts (a) and (b) of the figures relate to the subsets (a) and (b) of the second set of the parameters. The SMD results obtained using the power-4 FK, power-3 KSA, symmetric power-2 (SPW), asymmetric power-2 (APW), and MF closures are shown as bold black, green, magenta, thin black, and cyan curves, respectively. Note that similar dependencies for $g(\xi)$ were predicted for homogeneous regions earlier \cite{OmelKoz} when studying the BDDC model in the spatially inhomogeneous case within the KSA closure. The IBM simulations data are presented by the blue and red curves (we use the same color scheme as for Figs.~1 and 2).

\begin{figure}[t]
\centering
\includegraphics[width=0.56\textwidth]{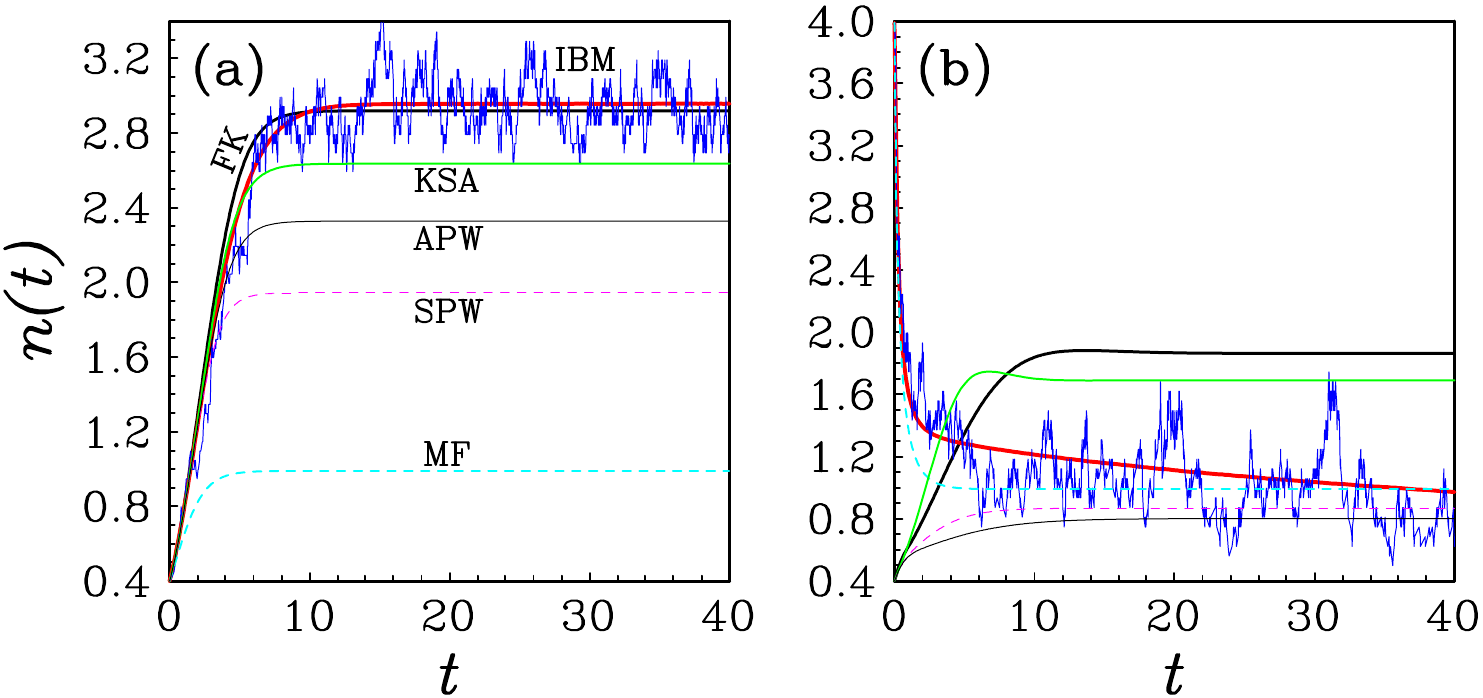}
\caption{First spatial moment (mean density) as a function of time obtained by the SMD approach for the BDDC model using the FK (bold black), KSA (green), SPW (magenta), APW (thin black), and MF (cyan) closures in comparison with that of the IBM simulations (red and blue). Parts (a) and (b) correspond to the two subsets of the model parameters from the second set.}
\label{f4}
\end{figure}

\begin{figure}[h]
\centering
\includegraphics[width=0.56\textwidth]{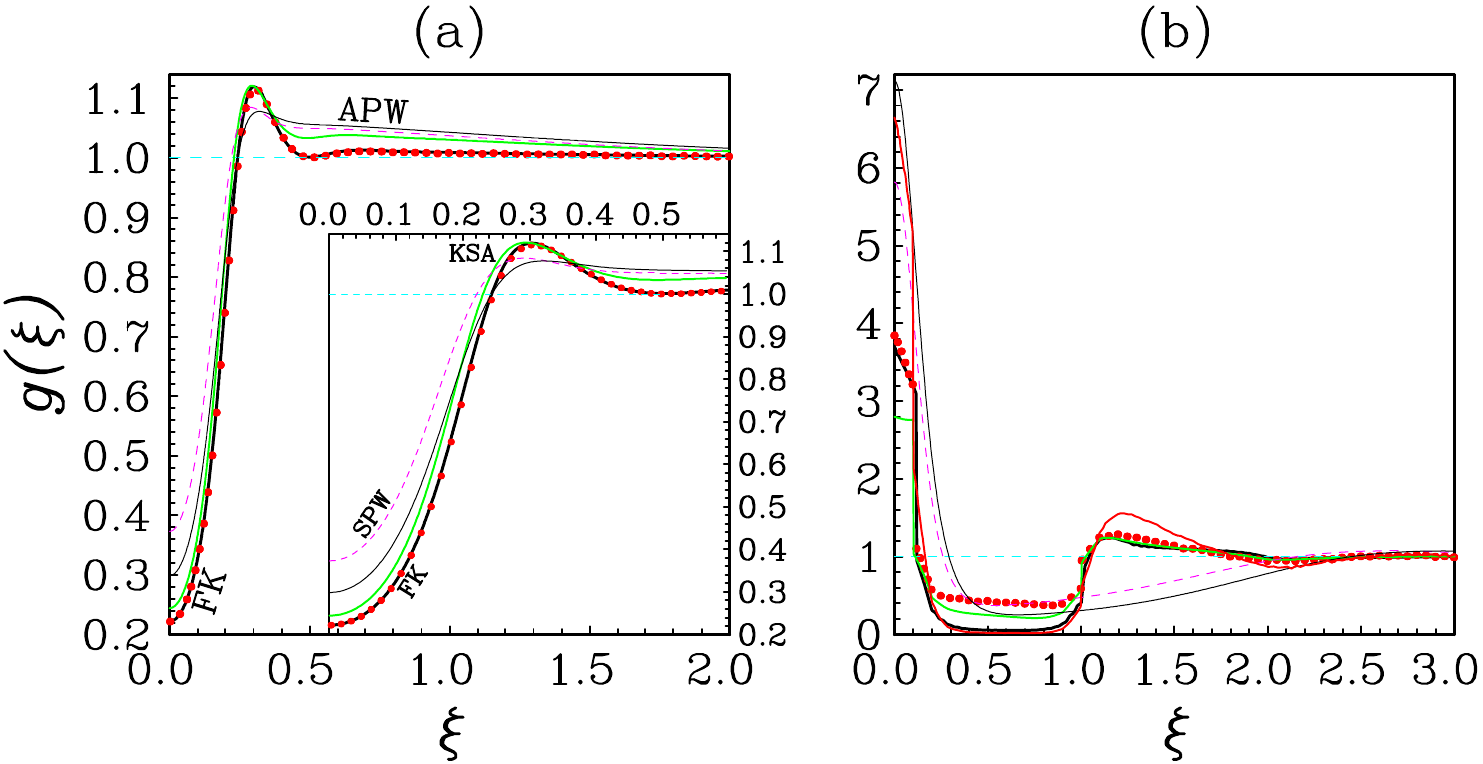}
\caption{Pair correlation function as depending on distance obtained by the SMD approach in the steady state using the FK (bold black), KSA (green), SPW (magenta), APW (thin black), and MF (cyan) closures in comparison with that of the IBM simulations (red circles). Parts (a) and (b) correspond to the same model subsets as those of Fig.~4, leading to deep segregation and extremely strong clustering, respectively.}
\label{f5}
\end{figure}

About differences between the closures for behaviour of $n(t)$ and $g(\xi)$ presented in Figs.~4(a) and 5(a) we can say nearly the same words as for those of Figs.~1(a) and 2(a). But now, when considering a deeply segregated state, these differences are much more visible. They decrease when arranging the closures in the following order: MF, SPW, APW, KSA, and FK. An evident convergence of the results to the ``exact'' IBM measurements is observed in Figs.~4(a) and 5(a) with increasing the hierarchical order of the SMD description. We see that the APW and SPW approaches, not to mention of MF, can not be used even for qualitative estimation of functions $n(t)$ and $g(\xi)$. Here the uncertainties can achieve more than 50\%. For instance, these approaches significantly underestimate density $n(t)$ in the steady state $t > \tau$ (which is achieved at $\tau \sim 20$). Moreover, they appreciably overestimate the depth of the segregation well, i.e. the value of $g(\xi)$ at $\xi=0$. In addition, the APW and SPW closures underestimate the maximal value of the first peak in $g(\xi)$ and incorrectly describe the long-tail asymptotic at larger distances. Better results are obtained within the KSA closure for which we can talk about a qualitative description. But the deviations are still apparent. Only when advancing to the FK ansatz we can say about a quantitative description in which the theoretical results are indistinguishable from the ``exact'' IBM data.

Take a look, finally, at the dependencies $n(t)$ and $g(\xi)$ in the deeply aggregated state. They are shown in Figs~4(b) and 5(b), respectively. Here the situation with the closures are quite different. All of them predict a steady-state behaviour of $n(t)$ at $t > \tau \sim 20$ with one or another value for $n_{\rm s}$. At the same time, the IBM function $n(t)$ continues to decrease even at $t = 40$, despite its initial value was chosen to be in ten times larger ($n(0) \approx 4$ versus 0.4) than that for the closure approaches SPW, ASPW, KSA, and FK. Because of this we extended the IBM simulations up to very long times $t=1000$ and the corresponding results are depicted in Fig.~6. From it we see clearly (red curve) that the density function (averaged over large number of ensembles) decays exponentially at long times, $\sim \exp(-t/\tau)$, with a relaxation time of $\tau \approx 300$ [the logarithmic scale was used in Fig.~6]. The time-dependent density related to a single realization (blue curve) exhibits huge fluctuations since the size of the system is finite ($\mathcal{L}=16$ for this subset). So nonzero density values cannot be smaller then $1/\mathcal{L}=0.0625$ because the minimal number of entities in the population is equal to 1. For a randomly chosen realization, the population ceased to exist (see vertical blue line) at $t \approx 900$ when the number of entities suddenly breaks to 0. Thus, at the parameters of the model that corresponds to the current subset, the population will eventually go extinct, i.e., $\lim_{t \to \infty} \mathcal{N}_t \to 0$ and $\lim_{t \to \infty} n(t) = 0$.

In view of Figs.~4(b) and 6(a), we can state that none of the closures are applicable for the last subset to describe properly the behaviour at long times. However, they can be exploited to estimate the pair correlations at short times, where the IBM density is comparable with its steady-state values in the SMD approach obtained by the closure approximations. Because of this, the pair correlation function $g(\xi,t)$ obtained in the IBM simulations is shown in Fig.~5(b) at $t=1$ (red circles) and $t=4$ (red curve) for comparison with the SPW, APW, KSA and FK results in the steady state. We can observe strong clustering of entities at small distances $\xi < s^\textrm{(B)}=0.1$, where $g \sim 4-7$, followed by their deep disaggregation in a wide range, $0.1=s^\textrm{(B)} < \xi < \sigma^\textrm{(D)}=1$, where $g \approx 0$, with further slight aggregation at longer separations, leading to the asymptotics behaviour $ g \to 1$ at $\xi >3$. All these features are quantitatively reproduced by the KSA and FK closures to a greater or lesser extent. The two discontinuities of $g(\xi)$ at $\xi=s^\textrm{(B)}=0.1$ and $\xi=\sigma^\textrm{(D)}=1$ are explained by the discontinuous character of the top-hat dispersal and competition kernels [Eqs.~(16)]. The SPW and APW closures are clearly inferior with respect to KSA and FK, especially in reproduction of the segregated wall and behaviour at intermediate and long distances.

Examples of the spatial structures, corresponding to the second spatial moments plotted in parts (a) and (b) of Fig.~5, are explicitly shown in parts (a) and (b) of Fig.~D1 (see Appendix D) using positions of entities taken from the IBM simulations at $t=40$ and $t=4$, respectively. Mention that these structures relate to the deeply segregated [part (a)] and strongly aggregated [part (b)] states. A spatial pattern of a Poisson process is also given there for comparison.

\begin{figure}
\centering
\includegraphics[width=0.56\textwidth]{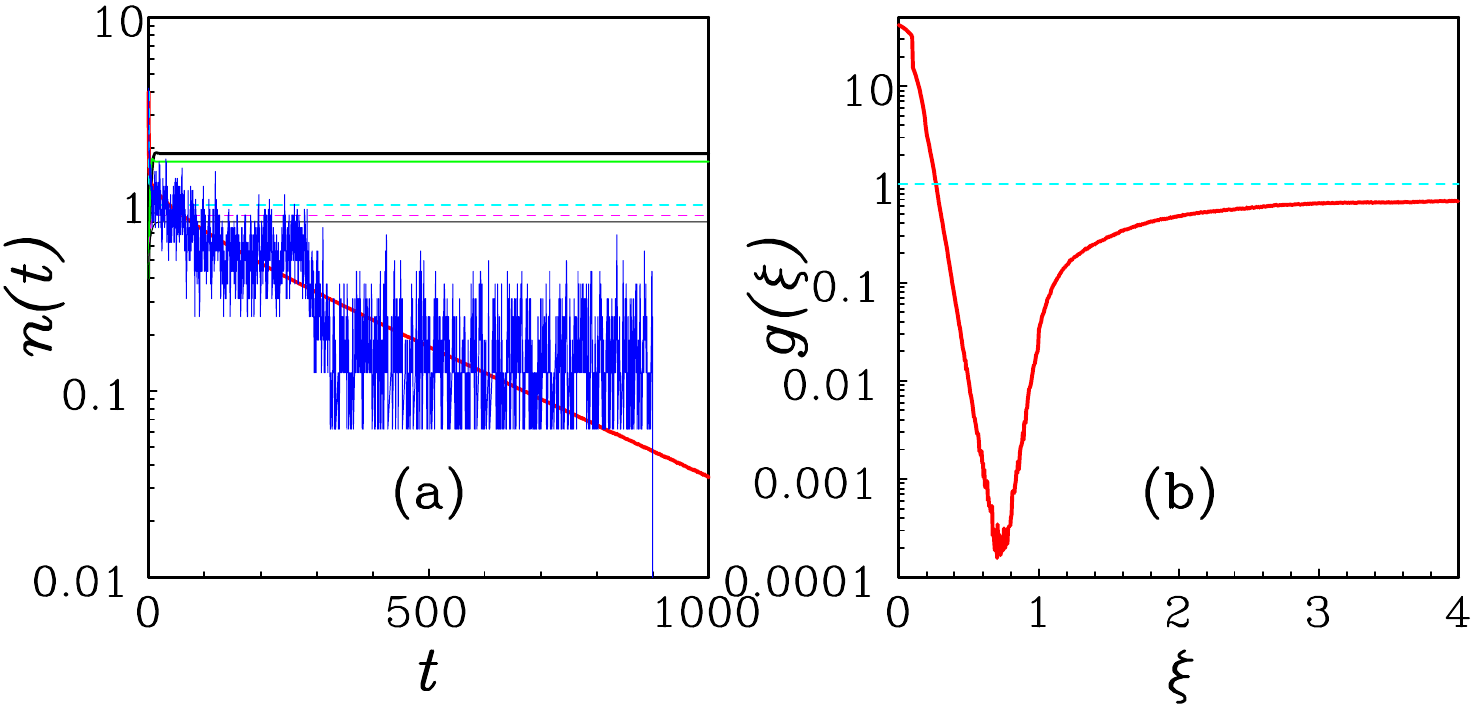}
\caption{First spatial moment [part (a)] and pair correlation function as depending on distance [part (b)] at long times in the case of extremely strong clustering followed by deep disagregation. Other notations are the same as in Figs.~4 and 5.}
\label{f6}
\end{figure}

The pair correlation function obtained in the IBM simulations by averaging over time in the interval $t \in 500-1000$ for the second subset is plotted in Fig.~6(b). It behaves similarly to that at small times [see Fig.~5(b)] but exhibits further, more stronger clustering, $g(\xi) \sim 40$, at small distances $\xi \sim 0$. A long-tail asymptotic behaviour can also be noticed. We see that even at sufficiently large separations $\xi \sim 4$, the function $g(\xi)$ still continues to approach its asymptotic value. At the same time, no such long tails exist in $g(\xi)$ for other parameters (see Figs.~2 and 5), where the pair correlations relatively quickly decay with increasing $\xi$, so that already at $\xi \lesssim 3$ or lesser separations they can be completely neglected.

\subsection{Third spatial moment (sets {\rm I} and {\rm II})}

The normalized third spatial moment $G(\xi,\xi') = F^\textrm{(3)}(\xi,\xi',t) / (F^\textrm{(1)}(t))^3$ (aka triplet correlation function) obtained by the SMD/FK approach for the four subsets of the first set (I) in the steady state ($t >\tau=20-200$) is shown as a surface in parts (a), (b), (c) and (d) of Fig.~7. The corresponding results for the two subsets of the second set (II) are displayed in parts (a) and (b) of Fig.~8 for the steady-state ($t >\tau=20$). In Fig.~8(b) we plotted $\ln G(\xi,\xi')$ for a better view. It can be seen that the dependencies of $G(\xi,\xi')$ on $\xi$ and $\xi'$ are similar to those of the pair correlation function $g(\xi)$ on $\xi$ in all the six subsets considered [cf. Figs.~2 and 5]. The areas near minimums of $G(\xi,\xi')$ in Figs.~7(a) and 8(a) are related to segregated configurations, $G < 1$, while the peaks in Figs.~7(b), (c), (d) and 8(b) indicate about clustering, $G > 1$. The surfaces are highly symmetrical because the triplet function possesses up sixth properties of symmetry, $F^\textrm{(3)}(\xi,\xi') = F^\textrm{(3)}(\xi',\xi) = F^\textrm{(3)}(-\xi,-\xi') = F^\textrm{(3)}(-\xi,\xi'-\xi) = F^\textrm{(3)}(\xi'-\xi,-\xi) = F^\textrm{(3)}(-\xi',\xi-\xi') = F^\textrm{(3)}(\xi-\xi',-\xi')$. The absolute minimum and maximum are achieved at $\xi=\xi'=0$ that corresponds to triplet configurations in which three entities are merged in one point of coordinate space. Intermediate extrema are exhibited at $\xi=0$, $\xi'=0$, or $\xi=\xi'$. They are responsible for situations when two of tree entities have the same positions. The asymptotic (MF) behaviour $G(\xi,\xi') \to 1$ at $|\xi| \to \infty$ or $|\xi'| \to \infty$ unless $\xi = \xi'$ as well as at $|\xi-\xi'| \to \infty$ unless $\xi'=0$ or $\xi=0$ is achieved at $|\xi|, |\xi'| > 0.2-2$.

\begin{figure}
\hspace{0.125\textwidth} (a) \hspace{0.5\textwidth} (b) \\
{\centering
\includegraphics[width=0.49\textwidth]{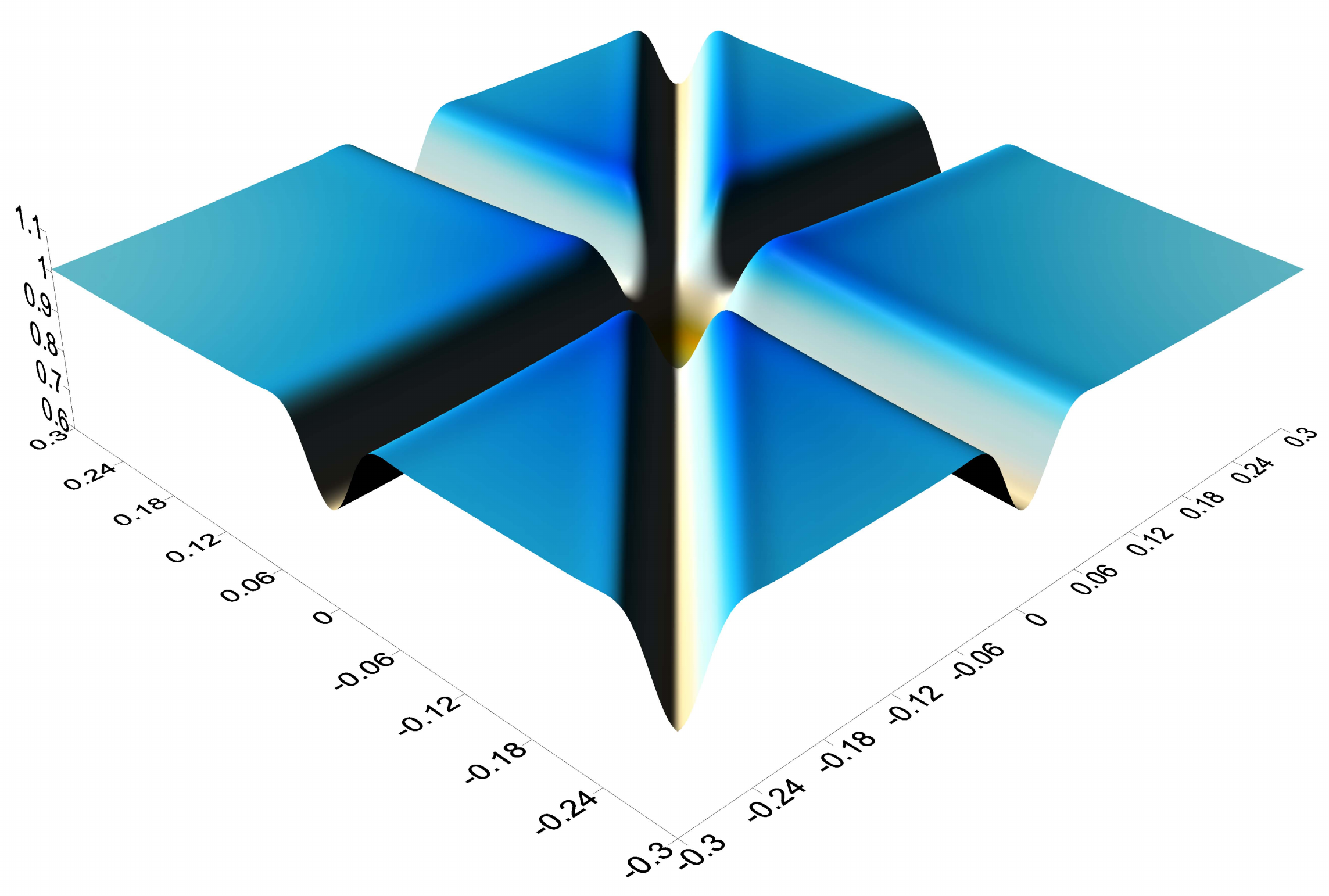}
\includegraphics[width=0.49\textwidth]{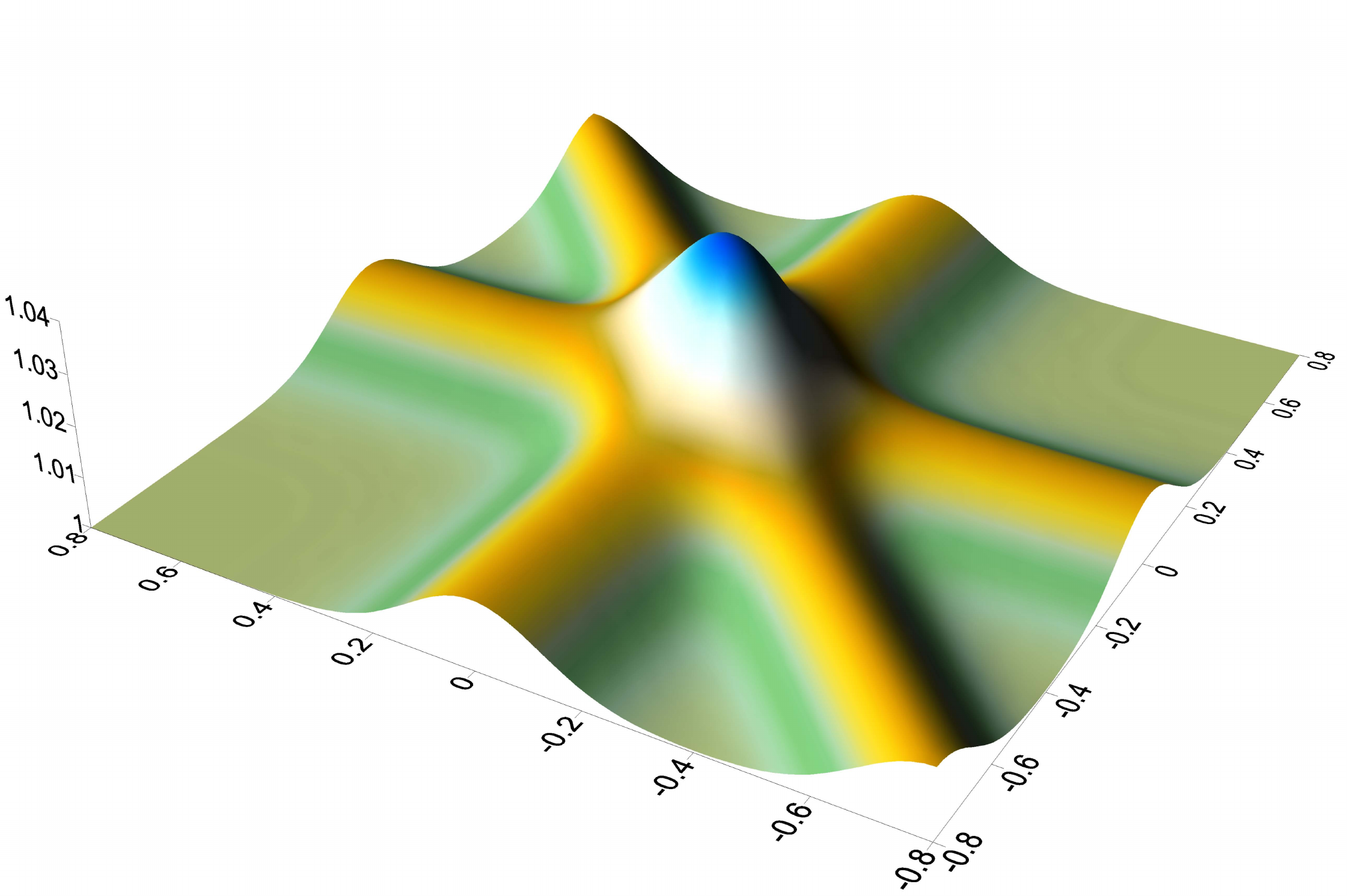}}
\hspace*{0.125\textwidth} (c) \hspace{0.5\textwidth} (d) \\
{\centering
\includegraphics[width=0.49\textwidth]{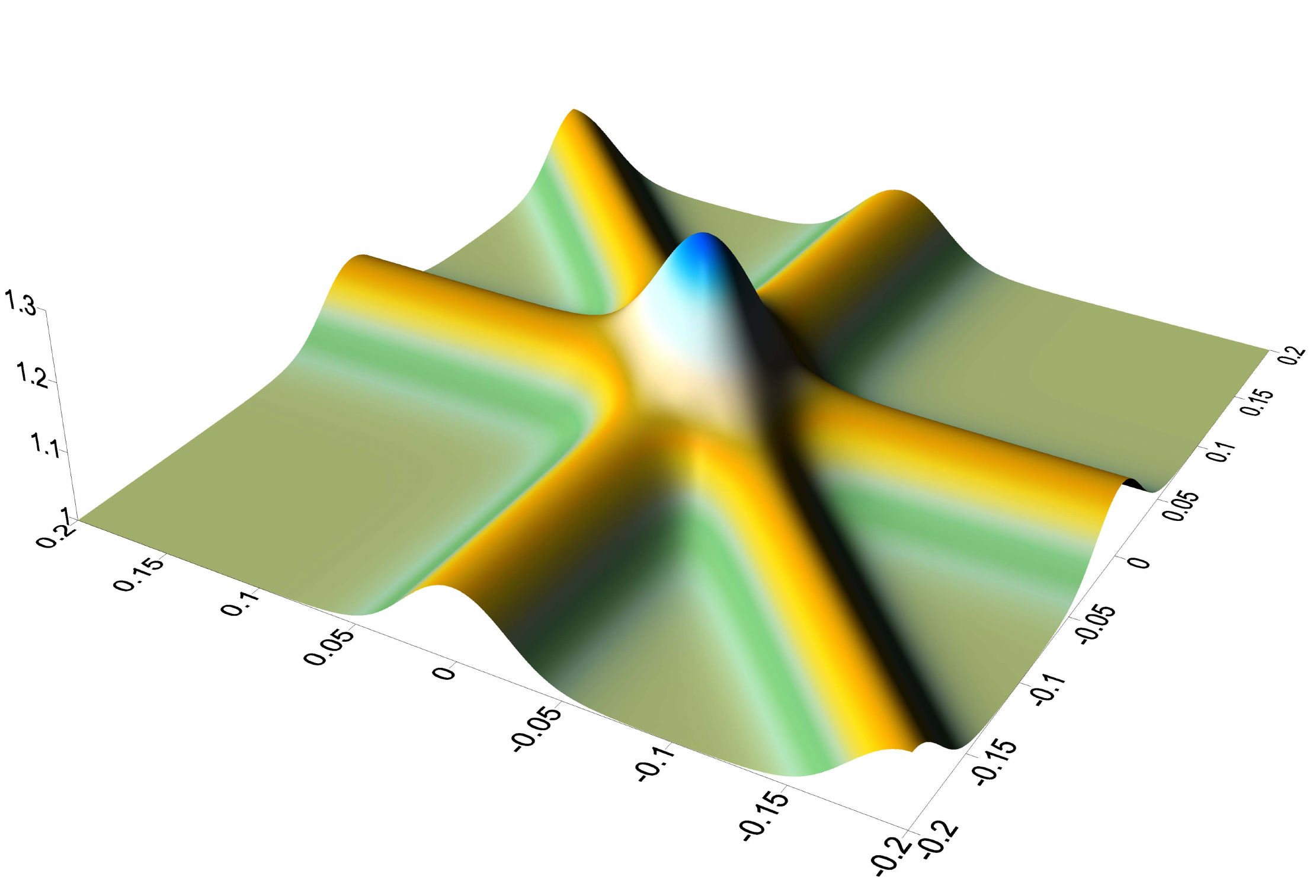}
\includegraphics[width=0.49\textwidth]{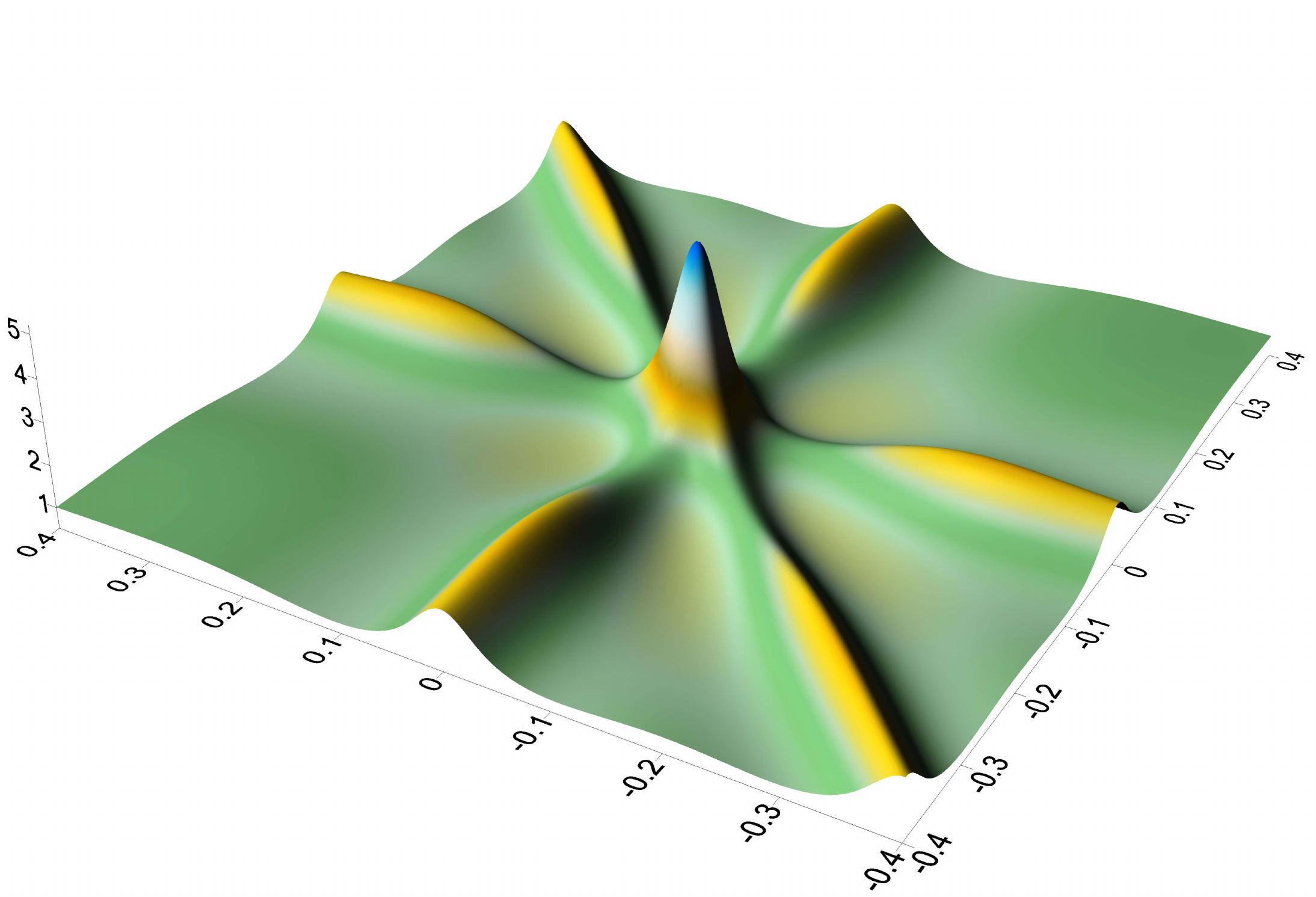}}
\caption{Triplet correlation function obtained by the SMD/FK approach for the four subsets [(a), (b), (c) and (d)] of the first set of the model parameters in the steady state. The subsets correspond to segregated, weakly, mildly aggregated, and strongly clustered spatial patterns, respectively.}
\label{f7}
\end{figure}

\begin{figure}
(a) \hspace{0.5\textwidth} (b) \\
\centering
\includegraphics[width=0.49\textwidth]{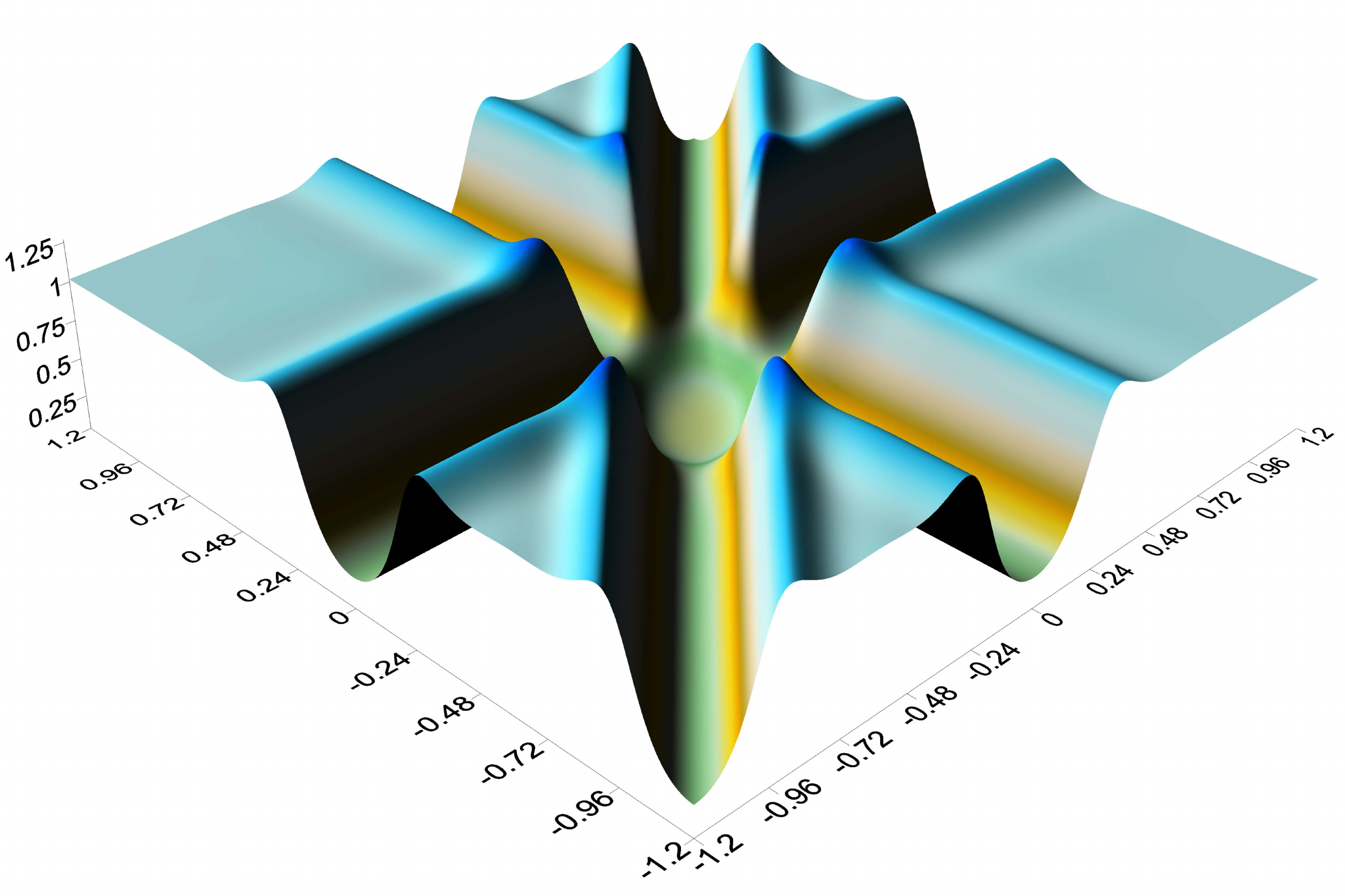}
\includegraphics[width=0.49\textwidth]{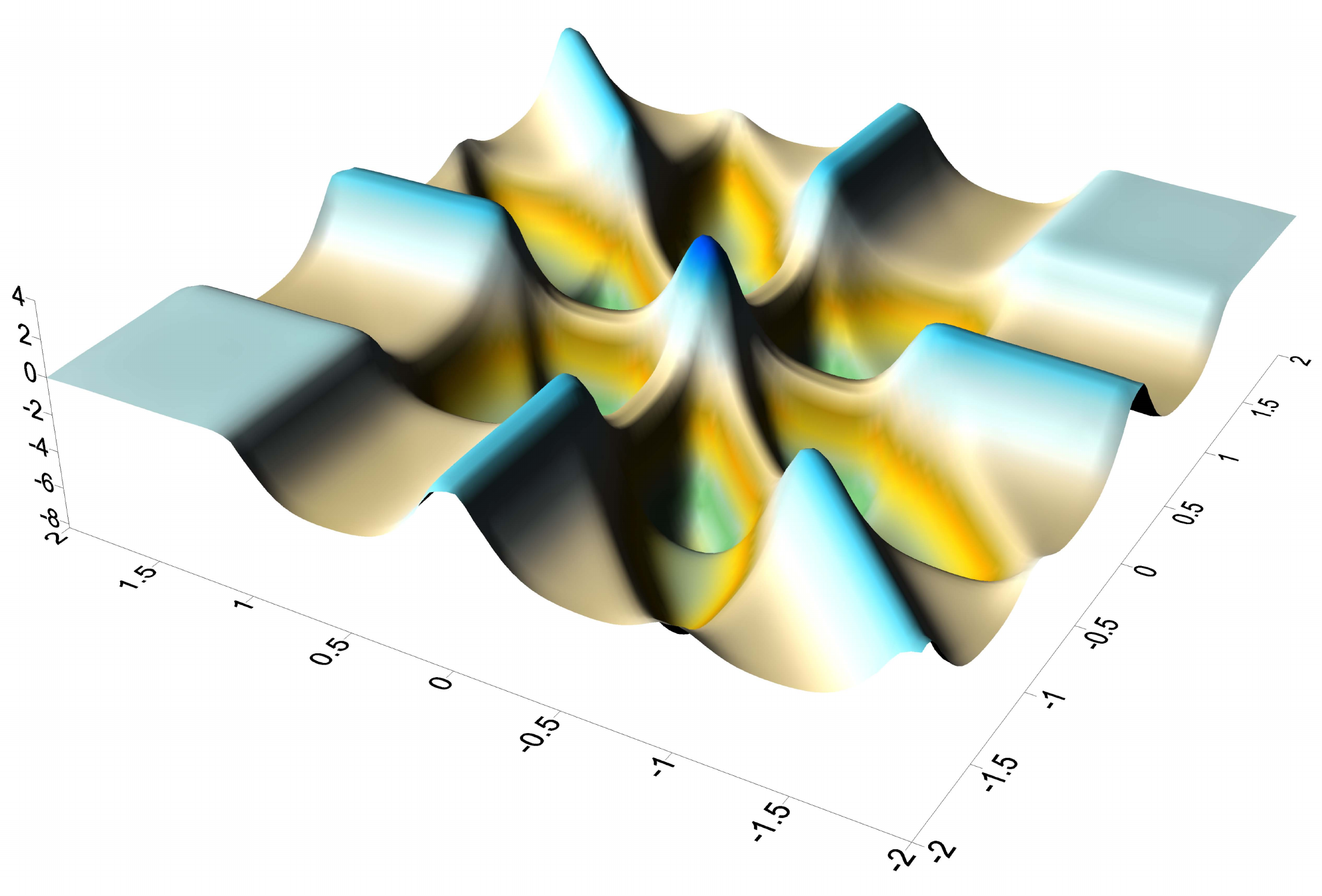}
\caption{Triplet correlation function obtained by the SMD/FK approach for the two subsets [(a) and (b)] of the second set of the model parameters in the steady state. The subsets correspond to deeply segregated and extremely strong aggregated spatial patterns, respectively.}
\label{f8}
\end{figure}

Remember that integrating the third-order correlation function on coordinate variables gives the averaged number of triplets in the system at a given time. Note also that the triplet distribution function contains information not only on the third-order dynamical spatial correlations in the system but on the second-order ones as well (see Appendix C). This can explain some similarity in dependence of $F^\textrm{(3)}(\xi,\xi')$ on $\xi$ and $\xi'$ with that of $F^\textrm{(2)}(\xi)$ on $\xi$. Moreover, it was shown (Appendix C) that for infinite systems in the homogeneous limit, the following equality takes place: $F^\textrm{(1)} F^\textrm{(2)}(\xi) = \lim_{L \to \infty} \frac{1}{2L} \int_{-L}^{L} F^\textrm{(3)}(\xi,\xi') {\rm d} \xi'$. From the latter equality it immediately follows that $\lim_{|\xi|,|\xi'| \to \infty} F^\textrm{(3)}(\xi,\xi') = F^\textrm{(1)} F^\textrm{(1)} F^\textrm{(1)}$, i.e., $\lim_{|\xi|,|\xi'| \to \infty} G(\xi) = 1$, where the relation $\lim_{\xi \to \infty} F^\textrm{(2)}(\xi) = F^\textrm{(1)} F^\textrm{(1)}$, i.e., $\lim_{\xi \to \infty} g(\xi) = 1$, has been taken into account (see Section {\em 4.3}). This confirms the MF asymptotic behaviour of $G(\xi,\xi') \to 1$ at large separation between all three entities in triplet configurations. Our estimations have shown that for $|\xi| \sim L$ and $|\xi'| \sim L$ the deviations of $g(\xi)$ and $G(\xi,\xi')$ from their asymptotic values did not exceed about $10^{-10}$ and $10^{-4}$, respectively, proving that the finite-size effects have been reduced to a minimum during the SMD/FK modeling. In the IBM simulations these deviations did not exceed a level ($\sim 10^{-4}$) of statistical noise.

\section{Conclusion}

An efficient numerical algorithm was proposed to solve the master equations for spatial moments of population dynamics on a higher level of description. It is based on a set of techniques which optimize the calculations by reducing the computational expenses to a minimum. This has allowed us to apply the superior Fisher-Kopeliovich (FK) closure of the fourth order to the hierarchy of the master equations in actual population dynamics simulations for the first time. As a consequence, the time-dependent population density, pair and triplet distribution functions were obtained within this closure for a birth-death model with nonlocal dispersal and competition in continuous space. By direct comparison with data retrieved from ``exact'' individual-based simulations as well as from the inferior mean-field, second-order and third-order Kirkwood superposition approximations, the evident superiority of the FK closure over the lower-order approximations has been demonstrated.

It can be concluded that the Fisher-Kopeliovich approach significantly improves the quality of the description in a wide range of varying parameters of the model. In particular, the FK ansatz is excellent for the study of any segregated and weakly aggregated states, where the theoretical results are indistinguishable from the ``exact'' data, leading to a quantitative description. This fourth-order anzatz is especially useful when investigating deeply segregated spatial patterns where all other known closures lead to the worst results. The FK approach is also good for moderately aggregated spatial structures, providing a qualitative reproduction. Enhancing by the next higher-order anzatz is required in the latter case to achieve quantitative description. Also, a modification of the FK closure is needed for strongly aggregated states.

The proposed SMD/FK approach can be applied to more complex SMD models \cite{Plank, Binny, Binnya, Surendran} which take into account motility, neighbour-dependent birth, different types, etc. The simulations can also be extended to higher dimensions. All of the above topics will be the subject of our future studies.

\section*{Appendix A}

\setcounter{equation}{0}
\renewcommand{\theequation}{A\arabic{equation}}

The first-, second-, and third-order spatial moments can be introduced as the macroscopic (coarse-grained) averages $\langle \ \rangle_{\rm m}$ of their microscopic counterparts followed by the ensemble (realization) averaging, giving expectation values \cite{Birch, Plank},
\begin{align}
F_a^\textrm{(1)}(x,t) =&\ \lim_{\mathcal{K} \to \infty} \frac{1}{\mathcal{K}} \sum_{\kappa=1}^{\mathcal{K}} \left \langle \sum_{\iota=1}^{\mathcal{N}_a^{(\kappa)}} \delta\big(x-x^{(\kappa)}_{a,\iota}(t)\big) \right \rangle_{\rm m} , \\
F_{ab}^\textrm{(2)}(x,y,t) =&\ \lim_{\mathcal{K} \to \infty} \frac{1}{\mathcal{K}} \sum_{\kappa=1}^{\mathcal{K}} \left \langle \sum_{a,\iota \ne b,\iota'}^{\mathcal{N}_{a,b}^{(\kappa)}} \delta\big(x-x^{(\kappa)}_{a,\iota}(t)\big) \delta\big(y-x^{(\kappa)}_{b,\iota'}(t)\big) \right \rangle_{\rm m} , \\
F_{abc}^\textrm{(3)}(x,y,z,t) =&\ \lim_{\mathcal{K} \to \infty} \frac{1}{\mathcal{K}} \sum_{\kappa=1}^{\mathcal{K}} \left \langle \sum_{a,\iota \ne b,\iota' \ne c,\iota''}^{\mathcal{N}_{a,b,c}^{(\kappa)}} \delta\big(x-x^{(\kappa)}_{a,\iota}(t)\big) \delta\big(y-x^{(\kappa)}_{b,\iota'}(t)\big) \delta\big(z-x^{(\kappa)}_{c,\iota''}(t)\big) \right \rangle_{\rm m} .
\end{align}
Here, $x^\textrm{($\kappa$)}_{a,\iota}(t)$ is the position of entity $\iota$ of type $a$ at time $t$ in the $\kappa$-th statistical ensemble (realization) with the total number $\mathcal{N}_a^\textrm{($\kappa$)}$ of $a$-type agents. The mean values of the moments are then obtained by summation over $\mathcal{K}$ independent realizations and division of the results on $\mathcal{K}$ in the infinite limit $\mathcal{K} \to \infty$. Each realization should correspond to the identical system but with a stochastically different initial ($t=0$) microscopic density distribution $\hat n_a^\textrm{($\kappa$)}(x,0) = \sum_\iota \delta(x - x^\textrm{($\kappa$)}_{a,\iota}(0))$ satisfying the condition $\lim_{\mathcal{K} \to \infty} \frac{1}{\mathcal{K}} \sum_{\kappa=1}^\mathcal{K} \langle \hat n_a^\textrm{($\kappa$)}(x,0) \rangle_{\rm m} = F_a^\textrm{(1)}(x,0)$. The same concerns $F^\textrm{(2)}(x,y,0)$ and $F^\textrm{(3)}(x,y,t,0)$.

The mean number of entities can be calculated as the ensemble averaging $\langle \mathcal{N}_a(t) \rangle = \lim_{\mathcal{K} \to \infty} \frac{1}{\mathcal{K}} \sum_{k=1}^\mathcal{K} \mathcal{N}_a^\textrm{($\kappa$)}(t)$. Contrary to the instantaneous integer number $\mathcal{N}_a^\textrm{($\kappa$)}(t)$ of agents of type $a$ at time $t$ in the $\kappa$-th realization, their mean value $\langle \mathcal{N}_a(t) \rangle \equiv \langle \mathcal{N}_a \rangle(t)$ is a continuous function of time. The spatial moments $F_a^\textrm{(1)}(x,t)$, $F_{ab}^\textrm{(2)}(x,y,t)$ and $F_{abc}^\textrm{(3)}(x,y,z,t)$ depend continuously on coordinates and time due to the macroscopic and ensemble averaging [Eqs.~(A1)--(A3)].

\section*{Appendix B}

\setcounter{equation}{0}
\renewcommand{\theequation}{B\arabic{equation}}

Let us prove that the APW closure does not satisfy the properties of symmetry for the third-order spatial moment. Indeed, taking into account its original form \cite{Law, Murrell},
\begin{equation}
F^\textrm{(3)}(\xi,\xi') = \frac{1}{\alpha+\gamma} \left[ \alpha \frac{F^\textrm{(2)}(\xi) F^\textrm{(2)}(\xi')}{F^\textrm{(1)}} + \beta \frac{F^\textrm{(2)}(\xi) F^\textrm{(2)}(\xi'-\xi)}{F^\textrm{(1)}} + \gamma \frac{F^\textrm{(2)}(\xi') F^\textrm{(2)}(\xi'-\xi)}{F^\textrm{(1)}} - \beta {F^\textrm{(1)}}^3 \right] ,
\end{equation}
it can be verified readily that for $\alpha \ne \beta = \gamma$ up four symmetrical properties are violated, namely, $F^\textrm{(3)}(\xi,\xi') \ne F^\textrm{(3)}(-\xi,\xi'-\xi) \ne F^\textrm{(3)}(\xi'-\xi,-\xi) \ne F^\textrm{(3)}(-\xi',\xi-\xi') = F^\textrm{(3)}(\xi-\xi',-\xi')$ while conserving only two equalities, $F^\textrm{(3)}(\xi,\xi') = F^\textrm{(3)}(\xi',\xi) = F^\textrm{(3)}(-\xi,-\xi')$. On the other hand, all six such properties are fulfilled by SPW ($\alpha=\beta=\gamma$). The same is related to the KSA which reads \cite{Dieckmann, Murrell, Binny, Binnya}:
\begin{equation}
F^{(3)}(\xi,\xi') = \frac{F^{(2)}(\xi) F^{(2)}(\xi') F^{(2)}(\xi'-\xi)}{F^{(1)} F^{(1)} F^{(1)}} \, .
\end{equation}
We omitted time variable $t$ for $F^\textrm{(1)}$, $F^\textrm{(2)}$ and $F^\textrm{(3)}$ in the above symmetrical relations, as well as in Eqs.~(B1) and (B2) for the sake of simplicity.

All the symmetrical properties are also satisfied exactly by definition in the case of the FK closure because then the third-order moment is not approximated at all, instead it is included explicitly into the master equations. Note that these properties follow from the invariance of the third moment $F^{(3)}$ with respect to permutation of agents between themselves in triplet configurations, because the agents of a given type are identical (interchangeable).
Failure of this symmetry destroys such an identity, leading to an unphysical distribution of agents in coordinate space.

Another disadvantage of the APW and SPW closures (B1) is that they have the potential to predict negative (unphysical) values for $F^\textrm{(3)}$ because of the presence of the $\beta$-term (with sign minus). Moreover, it is not always obvious which weighting constants ($\alpha$, $\beta$, $\gamma$) are most appropriate, so that their APW optimal values can depend on the parameters of the system. No such problems appear within the KSA and FK closures which are free of any adjustable parameters and always generate positive values for the spatial moments by definition.

\section*{Appendix C}

\setcounter{equation}{0}
\renewcommand{\theequation}{C\arabic{equation}}

Integration of the first, second, and third spatial moments on coordinate variables gives the averaged numbers of single entities, their pairs and triplets in the system, respectively, at a given time. For the spatially-inhomogeneous case, these numbers are determined from the relations $\int F^\textrm{(1)}(x,t) {\rm d} x = \langle \mathcal{N} \rangle$, $\int \int F^\textrm{(2)}(x,y,t) {\rm d} x {\rm d} y = \langle \mathcal{N} (\mathcal{N}-1) \rangle$, and $\int \int \int F^\textrm{(3)}(x,y,z,t) {\rm d} x {\rm d} y {\rm d} z = \langle \mathcal{N} (\mathcal{N}-1)(\mathcal{N}-2) \rangle$, where the integration is performed over the infinite continuous space $\mathbb{R}^d$. If the Dirichlet boundary conditions $\lim_{x,y,z \to \pm \infty} F^\textrm{(1,2,3)} = 0$ are assumed, the averaged numbers can accept finite values. Such conditions mean that the initial spatial distribution of entities was localized in a confined region and nonzero values of the spatial moments cannot approach the infinite boundaries ($\pm \infty$) during the propagation over a finite time $t$. In the spatially homogeneous case we deal with the finite, sufficiently large system of size $L$ and apply the periodic (toroidal) boundary conditions for IBMS simulation or the asymptotic boundary conditions for the SMD/FK approach (see Section {\em 3.2}). Then the integration over the finite space $x,y,z \in [-L,L]^d$ yields finite averaged numbers. The averaged densities of single entities, their pairs and triplets for finite spatially homogeneous systems can be calculated using variables $\xi=y-z$ and $\xi'=z-x$ as
\begin{align}
\frac{1}{2L} \int_{-L}^L F^\textrm{(1)} {\rm d} \xi =&\ F^\textrm{(1)} = \langle \mathcal{N} \rangle/(2L) \, , \\ \frac{1}{2L} \int_{-L}^L F^\textrm{(2)}(\xi) {\rm d} \xi =&\ \langle \mathcal{N} (\mathcal{N}-1) \rangle/(2L)^2 \, , \\ \frac{1}{(2L)^2} \int_{-L}^L \int_{-L}^L F^\textrm{(3)}(\xi,\xi') {\rm d} \xi {\rm d} \xi' =&\ \langle \mathcal{N} (\mathcal{N}-1)(\mathcal{N}-2) \rangle/(2L)^3 \, ,
\end{align}
where we put $d=1$ and omit $t$ to simplify notations. In the infinite-system limit $L \to \infty$, $\mathcal{N} \to \infty$ provided $\mathcal{N}/L \to \textrm{const}$ we come to the finite densities in the rhs of Eqs.~(C1)--(C3).

Taking into account Eqs.~(C1) and (C2), one obtains the integral relation between the first and second spatial moments:
\begin{equation}
F^\textrm{(1)} = \frac{\langle \mathcal{N} \rangle}{\langle \mathcal{N} (\mathcal{N}-1) \rangle} \int F^\textrm{(2)}(\xi) {\rm d} \xi \, .
\end{equation}
Analogously, in view of Eqs.~(C2) and (C3), one finds
\begin{equation}
F^\textrm{(2)}(\xi) = \frac{\langle \mathcal{N} (\mathcal{N}-1)\rangle}{\langle \mathcal{N} (\mathcal{N}-1)(\mathcal{N}-2) \rangle} \int F^\textrm{(3)}(\xi,\xi') {\rm d} \xi' ,
\end{equation}
meaning that the triplet distribution function contains information not only on the dynamical third-order spatial correlations in the system but on the second-order ones as well. Note, however, that for birth-death spatially inhomogeneous finite systems, where the number $\mathcal{N} \equiv \mathcal{N}(t)$ of entities is not constant in time and, thus, can fluctuate, the second spatial moments cannot be, in general, completely reproduced from the third-order correlation function. The reason is that the coefficient before integrand in the rhs of Eq.~(C5) is unknown in advance and require both the moments for its determining. The above reproduction is possible only in two cases: spatially inhomogeneous systems with finite constant numbers $\mathcal{N} \ge 3$ of particles, where $F^\textrm{(2)}(\xi) = \int F^\textrm{(3)}(\xi,\xi') {\rm d} \xi' / (\mathcal{N}-2)$, or infinite systems in the homogeneous limit with
\begin{equation}
\eta^\textrm{(3)} F^\textrm{(1)} F^\textrm{(2)}(\xi) = \lim_{L \to \infty} \frac{1}{2L} \int_{-L}^{L} F^\textrm{(3)}(\xi,\xi') {\rm d} \xi' \, , \ \ \ \ \ \ \ \ \eta^\textrm{(3)} = \frac{\langle \mathcal{N} (\mathcal{N}-1)(\mathcal{N}-2) \rangle}{\langle \mathcal{N} \rangle \langle \mathcal{N} (\mathcal{N}-1)\rangle} \, ,
\end{equation}
where the third-order fluctuations ratio $\lim_{\mathcal{N} \to \infty} \eta^\textrm{(3)} = 1$. The same concerns the reproduction of the first moment from the second-order correlation function using equation (C4). For the finite and infinite systems this equation transforms to $F^\textrm{(1)} = \int F^\textrm{(2)}(\xi) {\rm d} \xi / (\mathcal{N}-1)$ at $\mathcal{N} \ge 2$ and
\begin{equation}
\eta^\textrm{(2)} F^\textrm{(1)} F^\textrm{(1)} = \lim_{L \to \infty} \frac{1}{L} \int_0^L F^\textrm{(2)}(\xi) {\rm d} \xi \, , \ \ \ \ \ \ \ \ \eta^\textrm{(2)} = \frac{\langle \mathcal{N} (\mathcal{N}-1) \rangle}{\langle \mathcal{N} \rangle^2} \, ,
\end{equation}
respectively, where the second-order fluctuations ratio $\lim_{\mathcal{N} \to \infty} \eta^\textrm{(2)} = 1$. In our SMD/FK simulations the above fluctuations ratios were very close to 1, confirming that the sizes of the systems have been chosen sufficiently large.

\section*{Appendix D}

\setcounter{figure}{0}
\renewcommand{\thefigure}{D\arabic{figure}}

Snapshots of entity locations taken from our IBM simulations related to the two subsets of the second set of the BDDC model parameters are shown in parts (a) and (b) of Fig.~D1. These snapshots correspond to one realization chosen at random from many ensembles for each of the two simulations at $t=40$ and $t=4$, respectively. Since we deal with the one-dimensional case, the entities were presented by vertical bar lines, where their horizontal positions coincide with those of the individuals (an interval $[-8,8]$ was chosen for convenience). For the purpose of comparison, an initial ($t=0$) spatial configuration of a randomly chosen ensemble is also included [part (c) of Fig.~D1]. Because in the IBM simulations the entities were distributed at random on the very beginning, we obtain a Poisson spatial configuration in Fig.~D1(c) with $g(\xi) \equiv 1$. Note that all three patterns have the same mean density (first moment). In other words, the total number of the vertical bar lines in the parts (a), (b) and (c) is the same, but their second moment is quite different [take a look at parts (a) and (b) of Fig.~5, where $g(\xi)$ deviates significantly from 1].

\begin{figure}[t]
\centering
\includegraphics[width=0.88\textwidth]{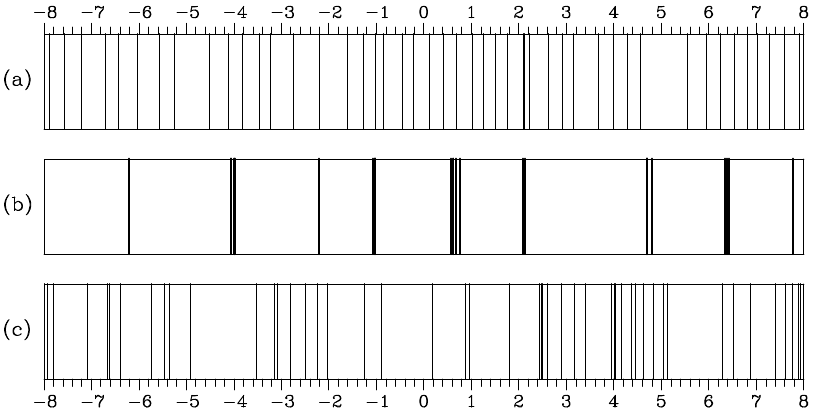}
\caption{Examples of spatial patterns corresponding to deeply segregated (a), strongly aggregated (b), and Poisson (c) states.}
\label{d1}
\end{figure}

We see that in the Poisson spatial pattern [Fig.~D1(c)] all agents’ locations are independent, i.e., the distribution of cells is completely random (like in systems of noninteracting particles). Here, the distance between agents can take arbitrary values with equal probability. On the other hand, Fig.~D1(a) describes a deeply segregated state related to the second moment shown in Fig.~5(a), where $g(0) \sim 0.2$ (effective repulsion between entities). For $g(\xi) < 1$, agent pairs separated by short distances are less likely to arise, generating a disaggregated regular spatial pattern (agents tend to be spaced apart). In contrast, Fig.~D1(b) represents a strongly aggregated state in which the second moment at small distances is large, see in Fig.~5(b), where $g(0) \sim 7$ (effective attraction at short separations). For $g(\xi) > 1$, pairs of cells are more likely to be found in close proximity than if they were distributed according to a Poisson pattern. Such a configuration corresponds to an aggregated spatial configuration where agents tend to be arranged in clusters (see many merged vertical bar lines, resulting in a few bold lines near which the entities coalesce).

\end{document}